\documentclass[linenumbers]{aastex631}

\usepackage{amsmath}
\usepackage{graphicx}
\usepackage{indentfirst}
\usepackage{booktabs}
\usepackage{subfigure}
\usepackage{url}
\usepackage{verbatim}
\begin{document}

\title{A 400Gbit Ethernet core enabling High Data Rate Streaming from FPGAs to Servers and GPUs in Radio Astronomy}

\author[0000-0002-9131-3664]{Wei Liu}
\affiliation{Department of Astronomy \\
 University of California, Berkeley \\
425 Campbell Hall \\
Berkeley, CA, 94720-3411, USA}

\author[0000-0003-1118-7731]{Mitchell C. Burnett}
\affiliation{Department of Electrical And Computer Engineering\\
Brigham Young University \\
Provo, UT, 84602, USA}

\author[0009-0008-7452-2027]{Dan Werthimer}
\affiliation{Department of Astronomy \\
 University of California, Berkeley \\
425 Campbell Hall \\
Berkeley, CA, 94720-3411, USA}

\author[0000-0003-0249-7586]{Jonathon Kocz}
\affiliation{Department of Astronomy \\
 University of California, Berkeley \\
425 Campbell Hall \\
Berkeley, CA, 94720-3411, USA}



\begin{abstract}

The increased bandwidth coupled with the large numbers of antennas of several new radio telescope arrays has resulted in an exponential increase in the amount of data that needs to be recorded and processed. In many cases, it is necessary to process this data in real time, as the raw data volumes are too high to be recorded and stored. Due to the ability of graphics processing units (GPUs) to process data in parallel, GPUs are increasingly used for data-intensive tasks. In most radio astronomy digital instrumentation (e.g. correlators for spectral imaging, beamforming, pulsar, fast radio burst and SETI searching), the processing power of modern GPUs is limited by the input/output data rate, not by the GPU's computation ability.  Techniques for streaming 
ultra-high-rate data to GPUs, such as those described in this paper, reduce the number of GPUs and servers needed, and make significant reductions in the cost, power consumption, size, and complexity of GPU based radio astronomy backends. In this research, we developed and tested several different techniques to stream data from network interface cards (NICs) to GPUs.  
 We also developed an open-source UDP/IPv4 400GbE wrapper for the AMD/Xilinx IP demonstrating high-speed data stream transfer from a field programmable gate array (FPGA) to GPU.
\end{abstract}



\section{Introduction}
Modern radio telescopes, such as the Square Kilometre Array (SKA)\cite{ska2007, ska2009}, the Atacama Large Millimeter/submillimeter Array (ALMA)\cite{alma2004}, Deep Synoptic Array(DSA)\cite{hallinan2019dsa, dsa20002022} and MeerKAT\cite{meerkat2016} generate data at unprecedented rates, reaching tens to hundreds of terabytes per second.  This data needs to be transported from data acquisition systems to real time processing units. There are different ways to process this data. Traditionally, an application specific integrated circuit based system, such as the Wideband Interferometric Digital ARchitecture(WIDAR)\cite{widar} correlator, would be used. Alternatives include FPGA and more recently FPGA/GPU based hybrid systems. \\

In radio astronomy, it is typically desirable to process as much frequency bandwidth as possible. High-speed ADCs are commonly used to process bandwidths of tens to hundreds of GHz\cite{adc2020}. To receive these data streams, FPGAs with high-speed transceivers are utilized to receive, channelize and then subsequently transfer them to data processing units\cite{fpga2016,fpga2021}. If these units are housed in a server (as is the case for CPU or GPU based processing), this is typically done via one or more NICs. Between the NICs and the FPGAs, high-speed Ethernet switch is also a critical part for the data transportation.\\

GPUs are renowned for their capacity for parallel processing, making them ideal for the computationally demanding jobs needed for astronomical data analysis. For instance, Fast Fourier Transforms (FFTs) are utilized in GPUs to transform time-domain signals into a frequency-domain representation\cite{gpufft2021}. The performance of GPU-based FFTs is typically much better than CPU-based FFTs\cite{fft-compare}. Correlators need to receive and process data from several stations. The most common correlator implementation is known as the "FX correlator", which first channelizes the band using a polyphase filter bank, and then cross-multiplies each antennas channelized data with every other antenna's data.  The computations can be done in parallel, which is ideal for GPUs\cite{xgpu2012}. In particular, Tensor-cores on the latest GPUs are special-purpose, matrix-matrix multiplication units that operate on limited-precision input data, which are good for correlator implementations\cite{gpucorrelator221}. Various other common elements of a data processing pipeline for radio telescopes (such an interference mitigation, calibration algorithms, and transient detection via beamforming and de-dispersion) can also be performed on GPUs \cite{gpurfi2017,gpufrb2020,gpufrb2024}.  \\

\begin{figure}[htbp]
	\centering
	\includegraphics[height=5cm]{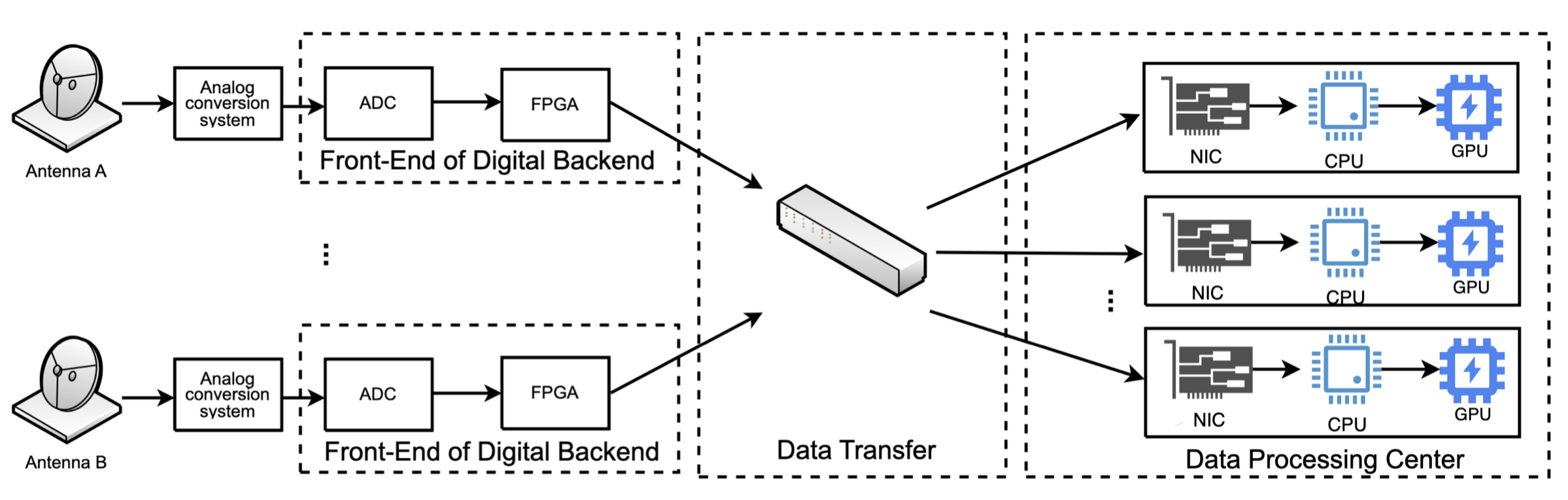}
	\caption{
		Diagram of a typical radio astronomy system showing the flow of data from telescopes to data processing centers. NICs, FPGAs, and GPUs are required to manage high-speed data streams.
		\label{radio-astronomy-system}
	}
\end{figure}

Modern GPUs are powerful enough for most radio astronomy data processing applications. However, moving data between different system components, such as FPGAs to NICs to GPUs is often a limiting step in the data processing pipeline. Utilizing high-throughput interconnects such as Peripheral Component Interconnect Express (PCIe) and NVIDIA's ``NVLink''\cite{nvlink2019} along with remote direct memory access (RDMA) technology\cite{rdma2016} can mitigate these issues, however, it is still difficult to design, develop and implement a high-performance data transportation system. Additionally, while there are existing solutions(such as GPUDirect) for transferring large amounts of data between two GPU servers, the initial ingest (from e.g. an FPGA based system) is typically done via UDP packets, and is not necessarily able to take advantage of these techniques. \\

The Collaboration for Astronomy Signal Processing and Electronics Research (CASPER) aims to reduce this hurdle through the development of open source, general purpose hardware, libraries, tools, and reference instrument
designs\cite{casper2016}. These tools enable researchers to implement the high-performance instrumentation needed to achieve their science goals in a cost effective and timely manner. \\

The adoption of 400GbE in radio astronomy offers significant improvements in data transmission and processing capabilities. As radio telescopes collect vast amounts of data, the need for faster, more efficient data handling becomes critical. 400GbE provides the necessary bandwidth to accommodate these large data rates, enabling real-time processing and analysis which are essential for timely discoveries and insights.\\

This paper outlines our research, beginning with the evaluation of RDMA techniques for bandwidth performance between NICs and GPUs. We then present the development of a 400GbE core with an FPGA for high-speed data transfer and an RDMA-based software framework for server-side data reception. This core will be integrated into the CASPER library, facilitating the rapid development of high-performance instrumentation by other researchers.\\

In Section~\ref{section-hardware}, the core hardware components used for this research are listed, and estimates the capability of the hardware established; The RDMA techniques for transporting high-speed data streams from NICs to GPUs are demonstrated in Section~\ref{section-technique}, which includes moving data through the host memory (Section~\ref{section-400g-nic-dram-gpu}), and using GPUDirect technology (Section~{\ref{section-400g-gpudirect}, \ref{section-400g-gpudirect-nic-2gpu}, \ref{section-400g-gpudirect-2nic-2gpu}); The implemented 400GbE core and associated test results for data transport from FPGA to NIC and then to GPUs are given in Section~\ref{section-400gbe-core-imp}; A summary of all results and outlines for future research directions are provided in Section~\ref{section-summary}.

\section{Hardware Requirements for 400GbE}
\label{section-hardware}
\subsection{System Description}
To transfer data from an FPGA to a GPU at 400 ~Gbps, it is important that all components in the data path are capable of operating at such speeds. This includes the FPGA, connecting cables, and all elements of the receiving server system. As RDMA is typically used to transfer high-speed data, RDMA technology should be supported by the NIC as well. A 400GbE NIC with a 16 lane PCIe 5.0 interface is required to support a theoretical maximum of 512 Gb/s between the NIC and other devices (DRAM controller, CPU or GPUs) on the PCIe 5.0 bus. A PCIe~5.0 motherboard and a PCIe~5.0 CPU are also necessary for the high data rate transportation. As DRAM is often used to transfer data to GPUs, having more DRAM controllers in the CPU will enhance performance. The NIC used for the tests described in this paper is the NVIDIA MCX75310AAS-NEAT\footnote{\url{https://docs.nvidia.com/networking/display/connectx7vpi/specifications}}, which supports PCIe~5.0 and 4x100~Gbps/lane; the CPU is Intel® Xeon® Silver 4410T\footnote{\url{https://www.intel.com/content/www/us/en/products/sku/232388/intel-xeon-silver-4410t-processor-26-25m-cache-2-70-ghz/specifications.html}}, which contains 80~lanes of PCIe~5.0 and eight DDR5 controllers; the motherboard is a Gigabyte MS03-CE0\footnote{\url{https://www.gigabyte.com/Enterprise/Server-Motherboard/MS03-CE0-rev-1x-3x}}, which has seven PCIe~5.0 slots.\\

To test packet transfers from the FPGA, we need to be able to implement a 400GbE core on the FPGA board, which requires high-speed transceivers at a minimum. Because the NIC supports 100Gbps/lane, the FPGA transceivers should also support 100Gbps/lane. We choose the VPK180\footnote{\url{https://www.xilinx.com/products/boards-and-kits/vpk180.html}} FPGA board based on AMD Versal Premium System on chip (SoC). Inside the Versal SoC, there are tens of GTM transceivers, which supports up to 112~Gbps when working in half-density mode. \\

The majority of GPUs currently on the market support up to PCIe~4.0,  capable of transfers of at least 200~Gbps. Therefore, we used two PCIe~4.0 GPUs for these tests. GPUs can be broadly classified into two categories: gaming GPUs and scientific, or compute, GPUs. In Section~\ref{section-technique} we compare the effectiveness of various data-moving techniques when comparing these two types of GPUs.
The GPUs we tested are the NVIDIA RTX A6000\footnote{\url{https://www.nvidia.com/en-us/design-visualization/rtx-a6000/}} and RTX 4070\footnote{\url{https://www.nvidia.com/en-us/geforce/graphics-cards/40-series/rtx-4070-family/}}. \\

Either copper cables or fiber optic links powered by transceivers are used for Ethernet interconnect. Inexpensive passive copper cables can be used for shorter distances (less than 3 meters for 400GbE), while fiber optic links, which can extend several kilometers, are needed for greater distances. In our work, we used copper cables from Fiberstore, which support the 400GBASE-CR4 interface standard. Table~\ref{core-hardware} summarizes the hardware used in this work.\\ 

 \begin{table}[htbp]
 \center
 \caption{Core Hardware for 400GbE Evaluation
  		\label{core-hardware}
  }
\begin{tabular}{|c|c|c|}
\hline
Hardware    & Part Number               & Specification                                                                             \\ \hline
CPU         & Intel® Xeon® Silver 4410T & \begin{tabular}[c]{@{}c@{}}80 lanes of PCIe 5.0;\\ 8 memory channels of DDR 5\end{tabular}       \\ \hline
NIC         & MCX75310AAS-NEAT          & \begin{tabular}[c]{@{}c@{}}PCIe 5.0;\\ 4 x 100Gbps/lane\end{tabular}           \\ \hline
Motherboard & MS03-CE0                 & \begin{tabular}[c]{@{}c@{}}7 x PCIe 5.0 slots;\\ 8-Channel DDR5 slots\end{tabular} \\ \hline
FPGA Board       & VPK180                    & Transceivers support up to 112Gbps                                                        \\ \hline
GPUs        & RTX A6000/ RTX 4070       & PCIe~4.0 scientific/gaming GPUs                                                            \\ \hline
Copper cables        & OSFP400-PC015       &  \begin{tabular}[c]{@{}c@{}}Support 400GBASE-CR4;\\ OSFP-RHS cable for the NIC to NIC test\end{tabular}                                                           \\ \hline
Copper cables        & OSFPFL-400G-QDDPC01       &  \begin{tabular}[c]{@{}c@{}}Support 400GBASE-CR4;\\ OSFP-RHS to QSFP-DD cable for the FPGA to NIC test\end{tabular}      
\\ \hline
\end{tabular}
\end{table}
 
\begin{figure}[htbp]
	\centering
	\includegraphics[height=6cm]{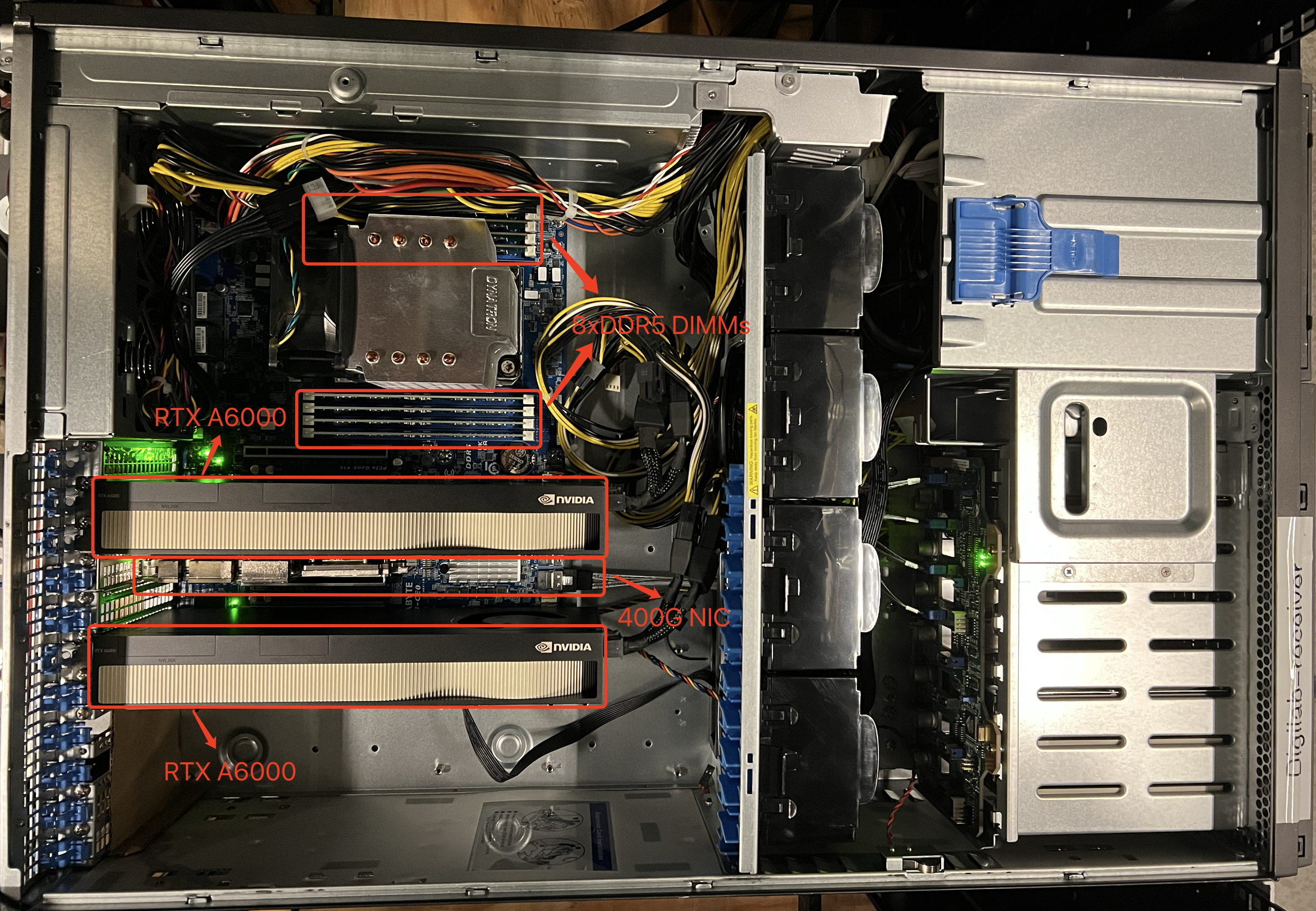}
	\caption{
		The server we setup for the 400G test, which includes 2 x RTX A6000(RTX 4070) GPUs, a 400G NIC, 8 x DDR5 DIMMs, a PCIe5.0 mother board and a PCIe5.0 CPU.
		\label{real-server}
	}
\end{figure}

\subsection{PCIe Bandwidth}
\label{PCIe-bandwidth}
PCIe is a high-speed bus interface standard for connecting the peripheral hardware components of a computer system. It serves as the critical technology enabling communication between the motherboard and peripherals, such as GPUs and NICs. The achievable bandwidth for the different PCIe versions is shown in Table~\ref{table-PCIe}. The above rates are unidirectional. Full-duplex bidirectional rates are twice the table values.\\

\begin{table}[h!]
\centering
\begin{tabular}{c c c c c}
\toprule
\begin{tabular}[c]{@{}l@{}}\textbf{PCIe}\\  \textbf{Version}\end{tabular}
 & \begin{tabular}[c]{@{}l@{}}\textbf{Release}\\  \textbf{Year}\end{tabular}
& \begin{tabular}[c]{@{}l@{}}\textbf{Transfer Rate}\\  \textbf{(GT/s per lane)}\end{tabular} & \begin{tabular}[c]{@{}l@{}}\textbf{Bandwidth}\\  \textbf{(GB/s per lane)}\end{tabular} & \begin{tabular}[c]{@{}l@{}}\textbf{Unidirectional  Bandwidth}\\  \textbf{(x16 configuration, GB/s)}\end{tabular} \\ 
\midrule
1.0 & 2003 & 2.5 & 0.250 & 4.0 \\ 
2.0 & 2007 & 5.0 & 0.500 & 8.0 \\ 
3.0 & 2010 & 8.0 & 0.985 & 15.75 \\ 
4.0 & 2017 & 16.0 & 1.969 & 31.5 \\ 
5.0 & 2019 & 32.0 & 3.938 & 63.0 \\ 
6.0 & 2022 & 64.0 & 7.877 & 126.0 \\ 
\bottomrule
\end{tabular}
\caption{PCIe Versions and Their Associated Data Transfer Rates}
\label{table-PCIe}
\end{table}

As the PCIe bus is the primary mechanism for data transfer used by majority of devices, including the NICs and GPUs, it is important to measure the PCIe bus bandwidth in order to estimate the maximum performance that is achievable. Table~\ref{core-hardware} indicates that the system we built for testing has a bottleneck on GPUs, which have a PCIe~4.0 interface, whereas other components support PCIe~5.0. To verify we achieve the PCIe 4.0 bandwidth, we used the code\footnote{\url{https://github.com/liuweiseu/bandwidthtest/blob/master/bandwidthtest.cu}} provided by NVIDIA to measure GPU bandwidth. The bandwidth performance for the RTX A6000 and RTX 4070 is shown in Figure~\ref{a6000-bandwidth} and Figure~\ref{4070-bandwidth}, respectively. Despite the substantial transfer rate achieved, this falls significantly short of the theoretical maximum transfer of 256~Gbps. The rates achieved do, however, match those achieved by NVIDIA with this code \footnote{\url{https://github.com/NVIDIA/nvbandwidth}}. As such, while it may be possible to achieve higher rates with, for example, other GPU models or drivers, we take this to be the limit of our current configuration for data capture and throughput testing purposes.\\ 

\begin{figure}[htbp]	
	\subfigure[RTX A6000 bandwidth performance]
	{
			\label{a6000-bandwidth}
			\includegraphics[height=4.5cm]{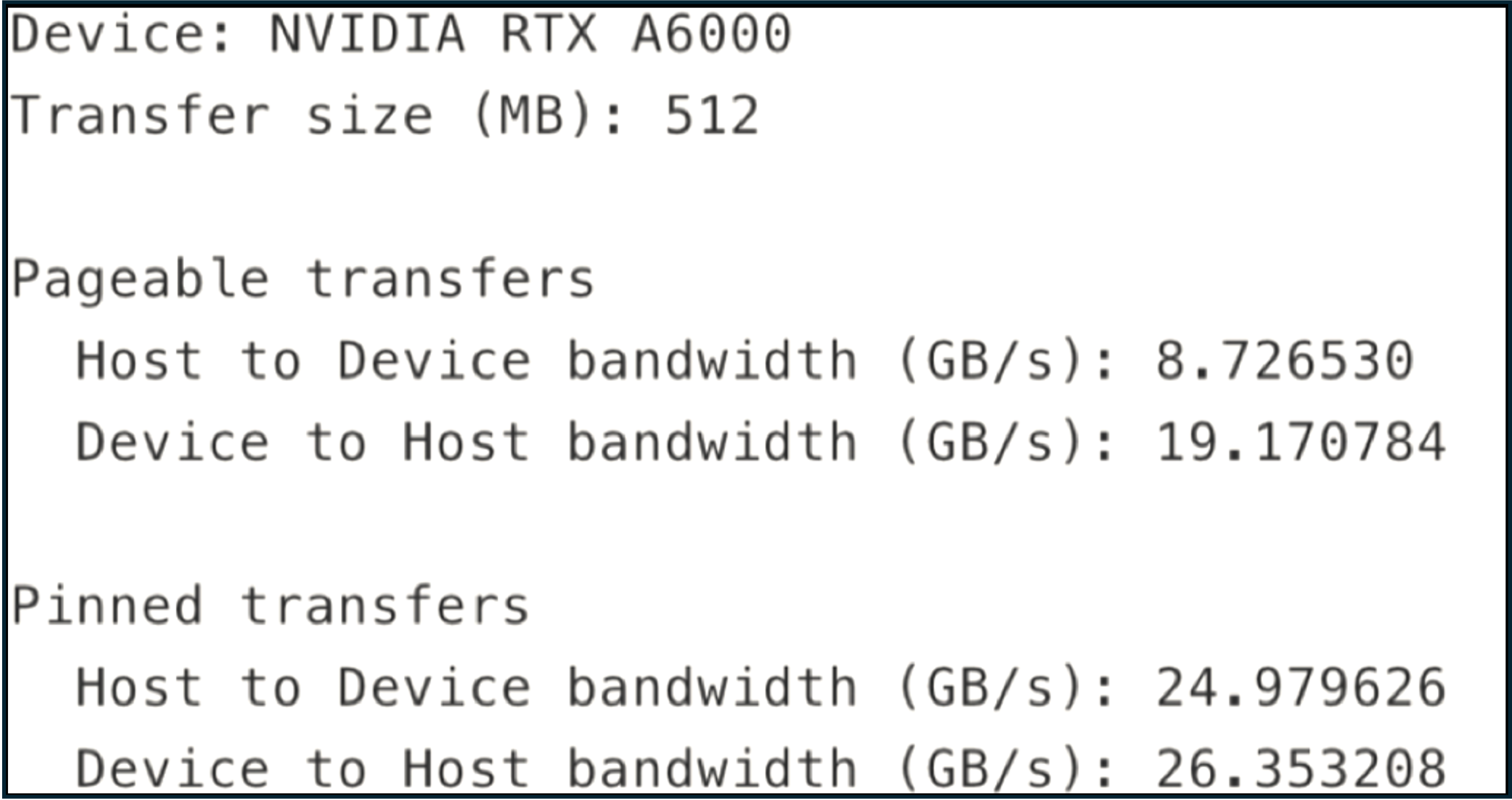}	}
	\subfigure[RTX 4070 bandwidth performance]
	{
			\label{4070-bandwidth}
			\includegraphics[height=4.5cm]{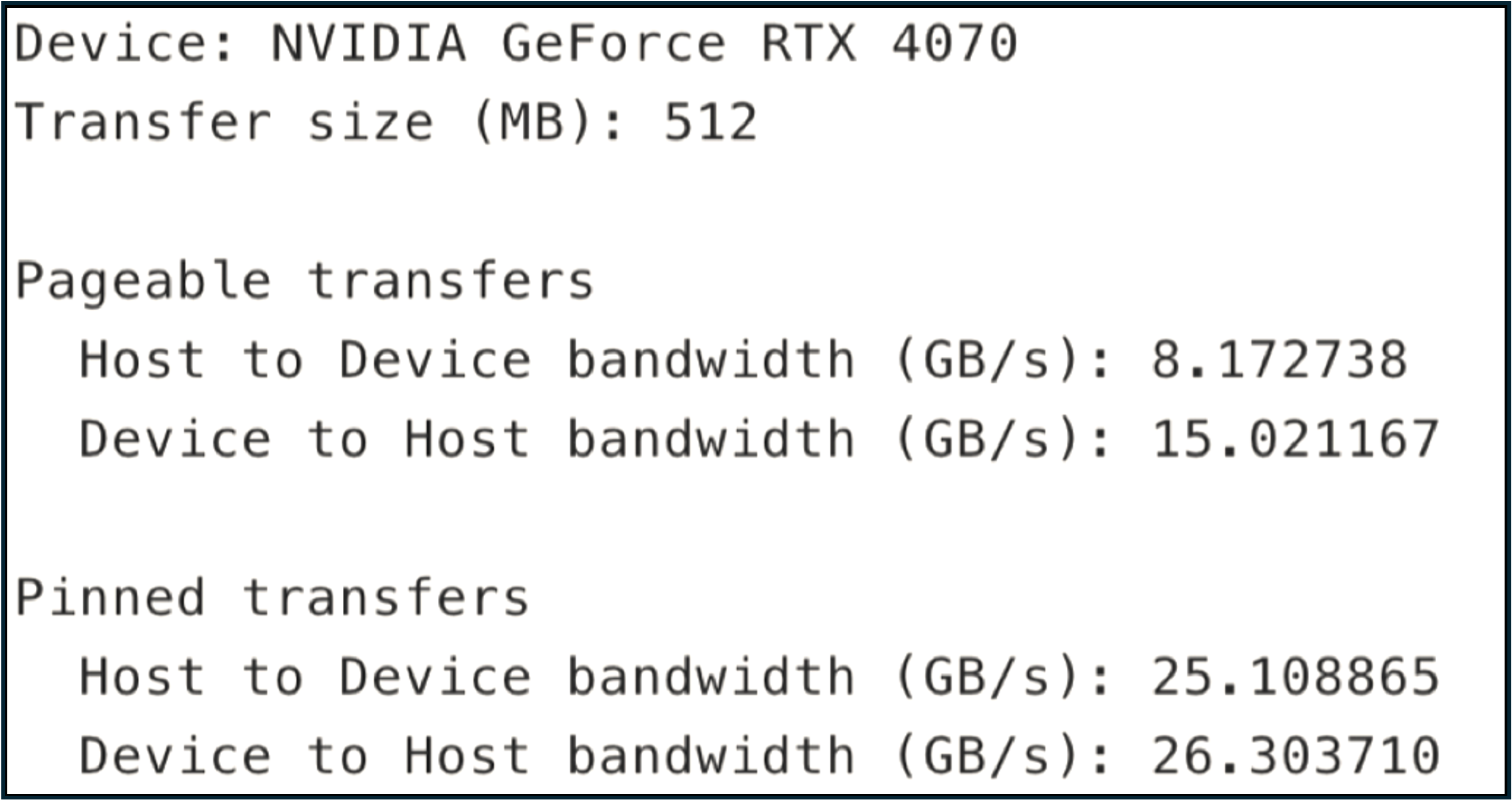}   
	}
	\caption{Measured bandwidth performance of the RTX A6000 and RTX 4070 GPUs over a PCIe 4.0 interface. The results indicate a maximum achievable bandwidth of $\sim$200~Gbps.} 
	\label{gpu-bandwidth} 
\end{figure}

 
 \subsection{Memory Bandwidth}
 \label{section-memory-bandwidth}
 High-speed data transfer also depends on memory bandwidth, which in this case must have a minimum capacity of 400~Gbps in order to accommodate simultaneous 400~Gbps data streams. With eight DDR5 DIMMs operating at 4800~MHz and an eight DDR5 controller built into the CPU, the memory bandwidth is sufficient. The memory bandwidth is measured using the Intel Performance Counter Monitor (Intel PCM) tool\footnote{\url{https://github.com/intel/pcm}} and stress-ng\footnote{\url{https://github.com/ColinIanKing/stress-ng}}\footnote{For best performance, please hand compile the benchmarks, since out of the box they wouldn't use the most efficient AVX.}, verifying that the 8-channel DDR5 memory bandwidth met the requirement for 400~Gbps data transfer (Figure~\ref{pcm-test-result}). \\

\begin{figure}[htbp]
	\centering
	\includegraphics[height=11cm]{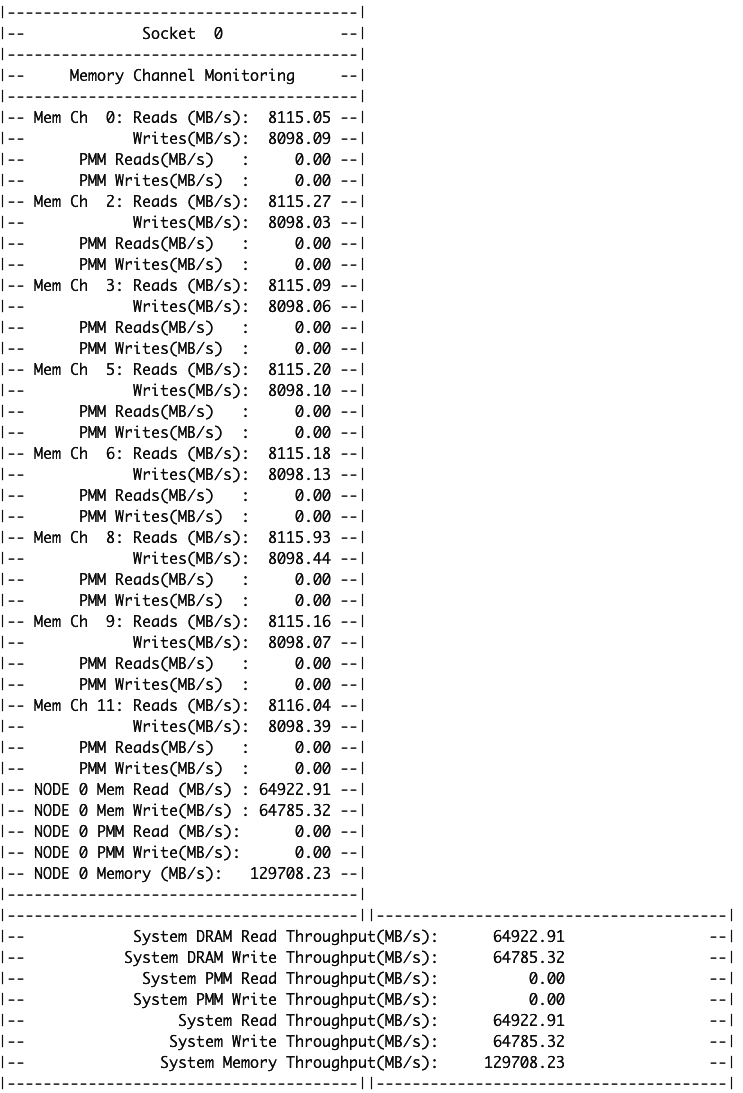}
	\caption{
		Memory bandwidth measurement of an 8-channel DDR5 memory configuration using the Intel Performance Counter Monitor (PCM) tool and stress-ng. The total memory bandwidth achieved is approximately $\sim$1013~Gbps, meeting the requirements for 400~Gbps data transfer.
		\label{pcm-test-result}
	}
\end{figure}

\section{Techniques for Data Transportation from NIC to GPU}
\label{section-technique}
Transmission Control Protocol/Internet Protocol (TCP/IP) and RDMA are crucial concepts for high-speed data transportation. Both techniques are used for data transfer, but they operate in fundamentally different ways and have distinct performance characteristics. This section discusses the differences between these approaches and showcases the performance of RDMA techniques using the test setup shown in Figure~\ref{400g-servers}. \\

\begin{figure}[htbp]
	\centering
	\includegraphics[height=4.5cm]{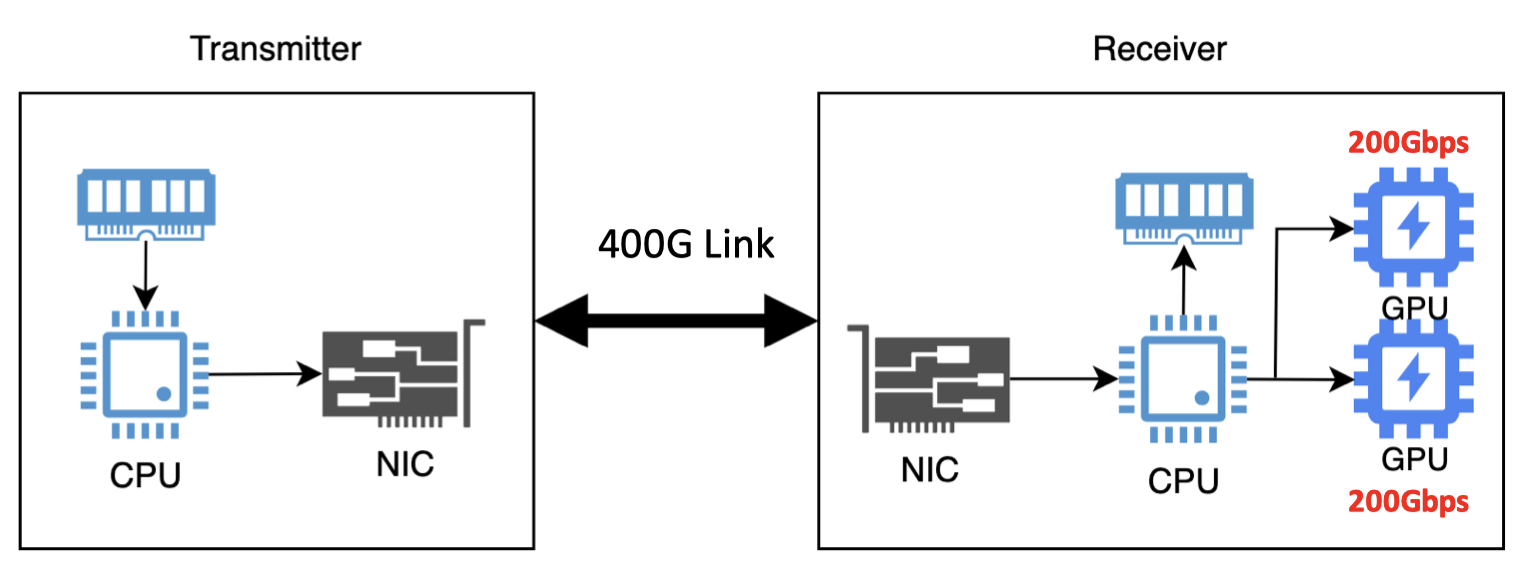}
	\caption{
		Setup of the 400G test servers used in the experiment. The servers are equipped with 400G NICs and GPUs to evaluate data transfer performance using various techniques.
		\label{400g-servers}
	}
\end{figure}

\subsection{Remote Direct Memory Access}
\label{section-tcpip}
TCP/IP, the fundamental set of communication protocols for the Internet and most local networks, consists of IP, TCP and UDP (User Datagram Protocol). Because of no buffering requirement, UDP is simpler to implement and allows for faster overall transmission speeds. In most radio astronomy applications, high-speed data transmission is in unidirectional, from FPGAs to a data processing center, through a dedicated network. Consequently, UDP is commonly implemented. \\

While it is simple to receive UDP streams using a Linux kernel UDP socket, it requires multiple buffer copies and heavy CPU usage, limiting its throughput. RDMA bypasses the host CPU and kernel, providing a high-throughput and low-latency network communication. Two outstanding features of RDMA are implemented in comparison to TCPIP: zero-copy and bypass kernel. The difference between UDP and RDMA is shown in Figure~\ref{rdma-introduction}. For a 400GbE network, RDMA technology is essential for optimal performance. \\

\begin{figure}[htbp]
	\centering
	\includegraphics[height=7cm]{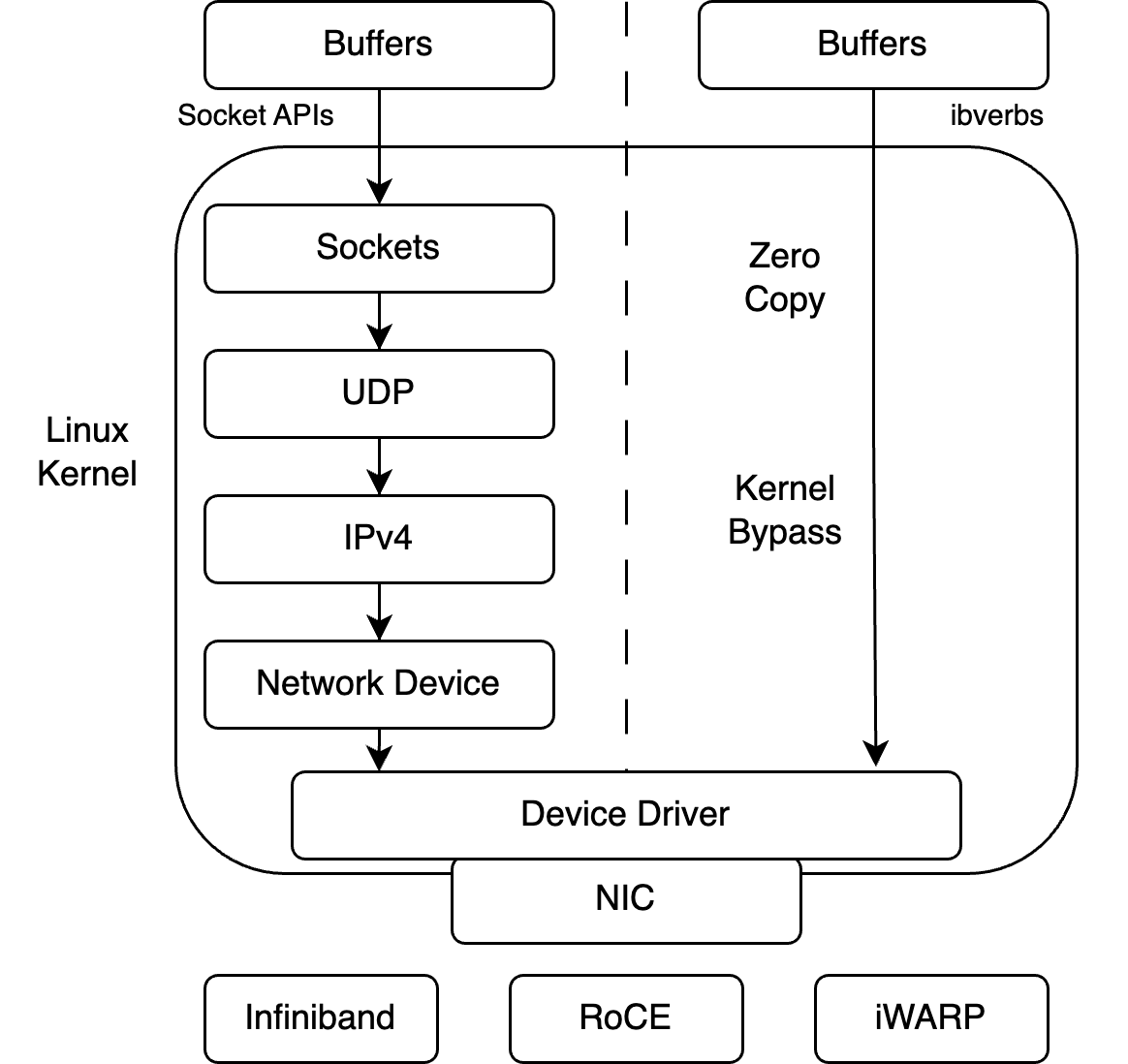}
	\caption{
		Comparison of data transfer processes using UDP and RDMA. The figure highlights the benefits of RDMA, including zero-copy and kernel bypass, which result in higher throughput and lower latency.
		\label{rdma-introduction}
	}
\end{figure}

Ibverbs (InfiniBand Verbs) is a low-level programming interface specifically made for InfiniBand/RDMA over Converged Ethernet(RoCE) networks that uses RDMA technology for high-performance network communication. Developers are able to design applications that demand exceptionally low latency and high throughput because to its direct access to the RDMA hardware. The OpenFabrics Enterprise Distribution (OFED) includes ibverbs as part of a broader suite of RDMA-supporting technologies. \\

For the 400GbE core test, we select RDMA with ibverbs based on performance, compatibility, complexity, and system upgradeability. The 400GbE core generates UDP packets. On the receiver side, the Queue Pair type created by ibverbs is RawEth, which captures UDP packets via kernel bypass and zero-copy.\\

\subsection{NIC to DRAM to GPU}
\label{section-400g-nic-dram-gpu}

Data transfers from NIC to GPU via DRAM is a general method compatible with both gaming and scientific GPUs. For 400GbE applications, a full-duplex memory bandwidth of at least 800~Gbps is required. Our servers meet this memory bandwidth requirement, as shown in Section~\ref{section-memory-bandwidth}. Figure~\ref{nic-dram-gpu} illustrates the data transfer path from NIC to GPU through DRAM.\\

\begin{figure}[htbp]
	\centering
	\includegraphics[height=4.5cm]{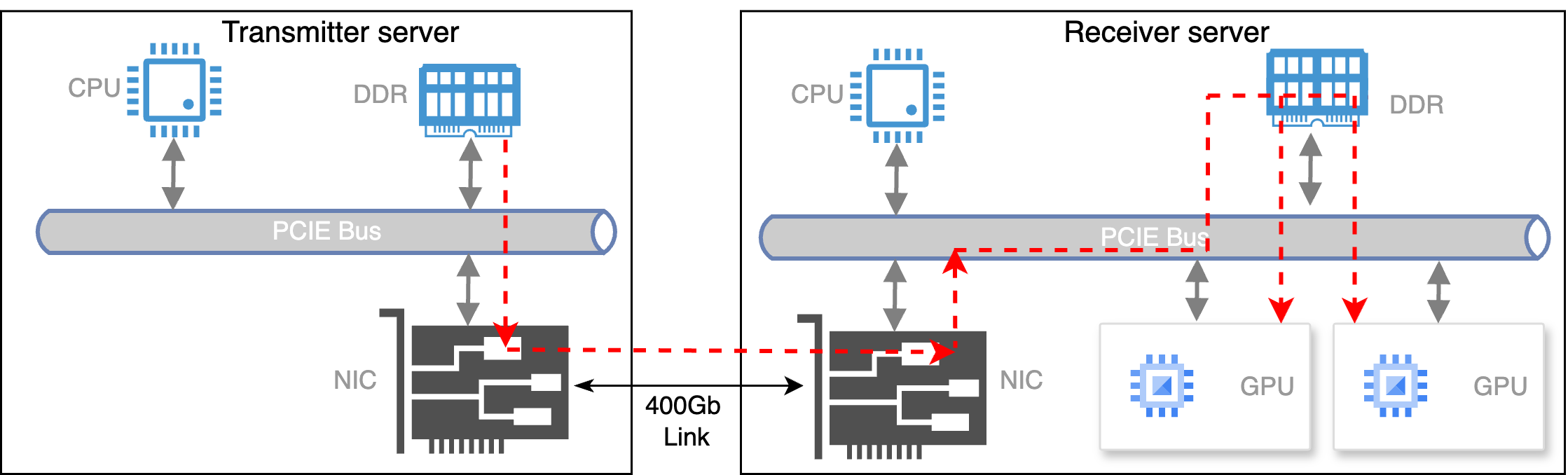}
	\caption{
		Data transfer path from NIC to GPUs via DRAM. This method utilizes the available memory bandwidth to handle high-speed data streams. Red dashed lines show the direction of data transfer.
		\label{nic-dram-gpu}
	}
\end{figure}

In order to estimate the efficiency of data transfer from a NIC to GPUs via DRAM, a packet generator is developed on the transmitting server that sends packets using a 400GbE link. The maximum data rate of the packet generator is $\sim$380~Gbps. As this exceeds the capture code capability, the speed of the generator is artificially limited to ensure we can the maximum data rate without packet loss. Each packet payload contains a packet sequence counter that was incremented by one. By examining the packet count sequence on the receiving end, we determine if there is any packet loss. \\ 

The open Source HASHPIPE framework\footnote{\url{https://github.com/david-macmahon/hashpipe}}\footnote{\url{https://github.com/liuweiseu/hashpipe-ibverbs-demo.git}}\cite{hashpipe2018} is used on the receiving server to capture packets and verify packet sequence values. Two threads, the network thread and gpu thread, are created in HASHPIPE for receiving packets and moving packets to the GPUs. A shared memory buffer of status values is provided by HASHPIPE for real-time status monitoring. The HASHPIPE framework used in the test is shown in Figure~\ref{hashpipe-nic-dram-gpu}.\\ 

\begin{figure}[htbp]
	\centering
	\includegraphics[height=4.5cm]{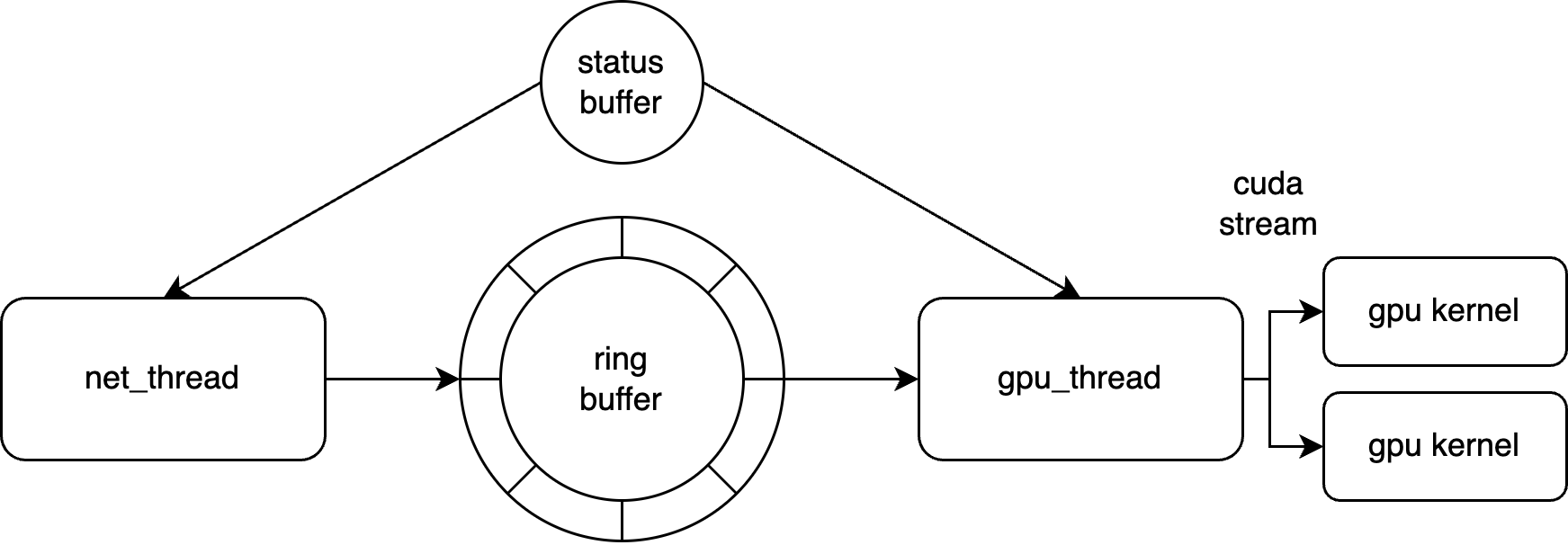}
	\caption{
		Architecture of the HASHPIPE framework used in the data transfer tests. The framework provides separate network and GPU threads for efficient packet handling, as well as provides real-time thread status monitoring.
		\label{hashpipe-nic-dram-gpu}
	}
\end{figure}

The ibverbs APIs are used in the network thread and allow for receiving packets at a high data rate. Once the network thread receive packets from the 400G NIC, they transport the data to shared ring buffers in the host RAM. Data from the ring buffer were transferred to two GPUs using two CUDA streams in the gpu thread. To determine the data transfer rate, we measure the amount of time it takes to fill the ring buffer. Additionally, we verify the counter sequence values in received packets checking for packet loss. The calculated data rate and packet loss are stored in real time using the status buffer, shown in Figure~\ref{a6000-nic-dram-gpu}(RTX A6000) and Figure~\ref{4070-nic-dram-gpu}(RTX 4070). \\

\begin{figure}[htbp]	
	\subfigure[RTX A6000 bandwidth performance]
	{
			\label{a6000-nic-dram-gpu}
			\includegraphics[height=3.4cm]{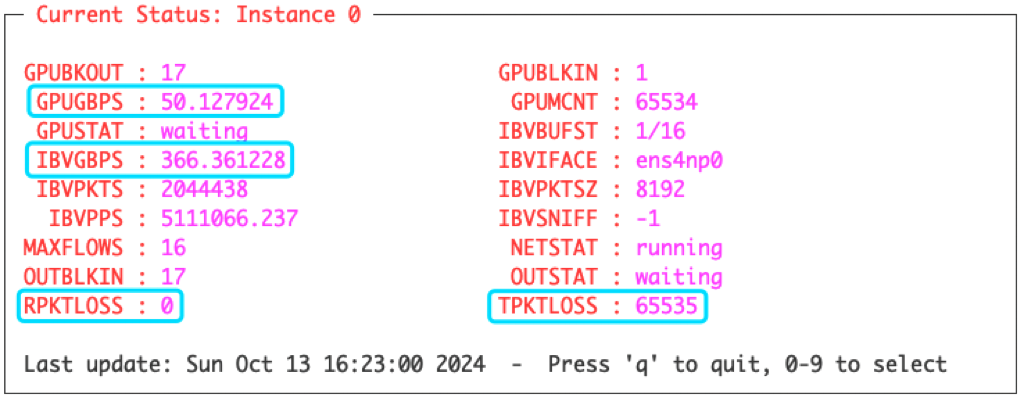}	}
	\subfigure[RTX 4070 bandwidth performance]
	{
			\label{4070-nic-dram-gpu}
			\includegraphics[height=3.4cm]{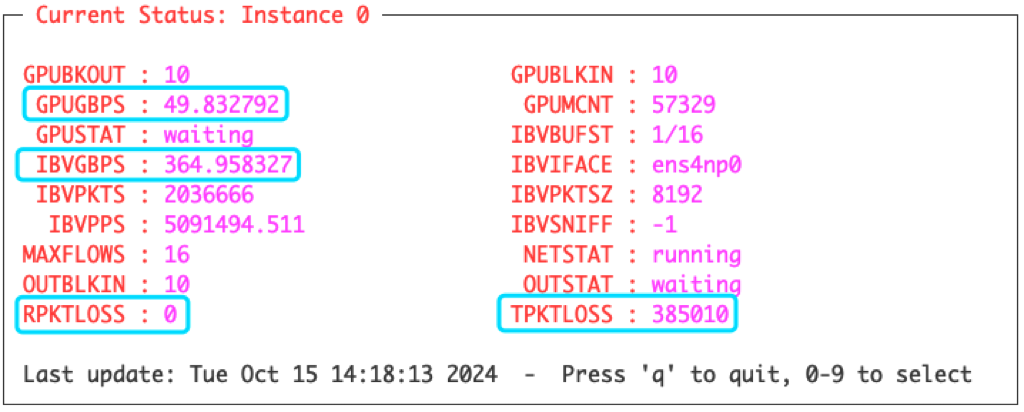}   
	}
	\caption{Measured bandwidth performance of the RTX A6000 and RTX 4070 GPUs during the NIC to GPUs via DRAM test. Results show that the data transfer rate without packet loss is about 360~Gbps. RPKTLOSS shows the real-time packet loss, which is always zero after the initialization; TPKTLOSS shows the total number of packet loss, which is non-zero during the initialization, but remains constant after initialization.} 
	\label{gpu-bandwidth-nic-dram-gpu}
\end{figure}

\newpage
\subsection{Direct Data Transport from NIC to a Single GPU}
\label{section-400g-gpudirect}

GPUs and NICs, both PCIe devices, benefit from PCIe's high bandwidth and low latency when RDMA is used. In order to directly transfer data from NICs to GPUs, specific NICs and GPUs are needed:  The NICs and GPUs should both be able to support RDMA. As we use NVIDIA NICs as well NVIDIA GPUs in this test, GPUDirect technology\footnote{\url{https://docs.nvidia.com/cuda/gpudirect-rdma/}} can be used. GPUDirect allows GPUs to communicate directly with other system parts, bypassing the CPU and system memory, reducing latency and increasing throughput.  Since both Tesla and Quadro GPUs support GPUDirect, we utilized the RTX A6000 for the GPUDirect test. Figure\ref{nic-to-gpu} shows the direct data transfer path from NIC to GPU using GPUDirect. As the PCIe limited IO bandwidth on the GPU is 200Gbps, we initially tested GPUDirect to one GPU with a 200G link. \\

\begin{figure}[htbp]
	\centering
	\includegraphics[height=4cm]{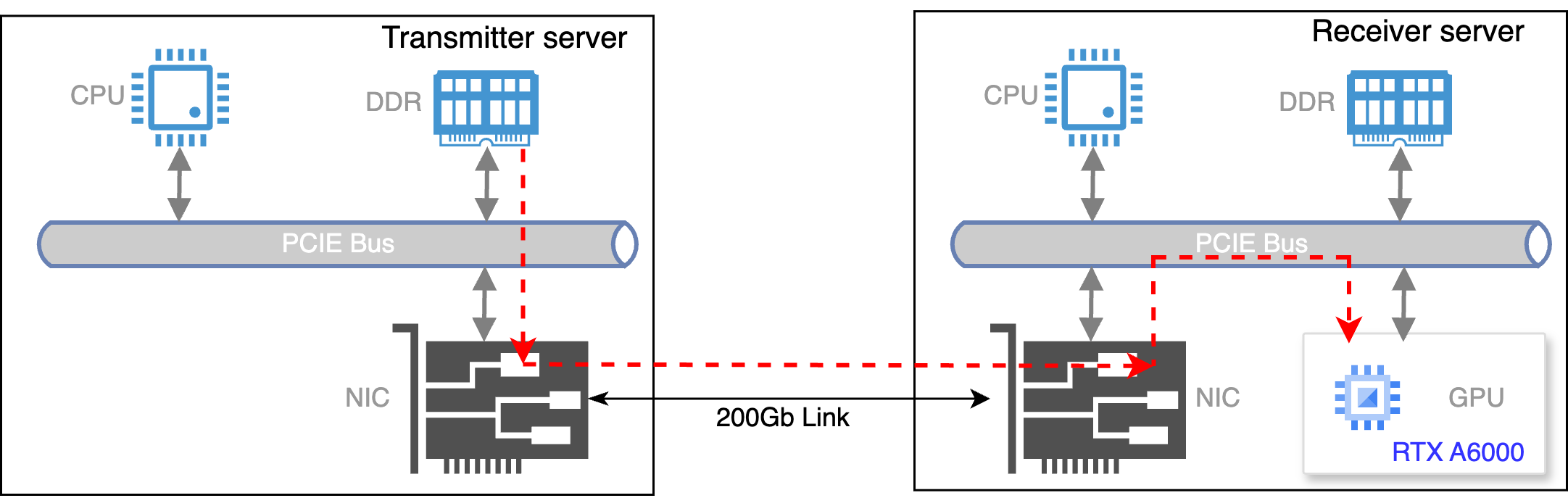}
	\caption{
		Direct data transfer path from NIC to GPU using GPUDirect technology. RDMA transactions reduces latency by bypassing the CPU and system memory. The link between the two servers is 200G. Red dashed lines shows the direction of data transfer.)
		\label{nic-to-gpu}
	}
\end{figure}
  
Packets on the transmitter server are generated using the ibverbs APIs and sent out in bursts to gauge bandwidth performance. Each burst transfers a total of 12.5~GB data, with each packet sizes to 8192 bytes and a total of 1638400 packets. The receiving server also uses ibverbs to capture and report packet counts. The ibverbs routine (ibv\_poll\_cq\footnote{\url{https://www.rdmamojo.com/2013/02/15/ibv_poll_cq/}}) reports how many packets are received successfully, so the packet loss can be checked by comparing the number of sent packets and the number of received packets. The RTX A6000 GPU supports PCIe~4.0, limiting throughput to 200Gbps. We achieved a maximum data rate of 178Gbps\footnote{\url{https://github.com/liuweiseu/400GbE_Demo.git}} without packet loss, as shown in Figure\ref{gpudirect-test-result}. \\

 \begin{figure}[htbp]
	\centering
	\includegraphics[height=5.8cm]{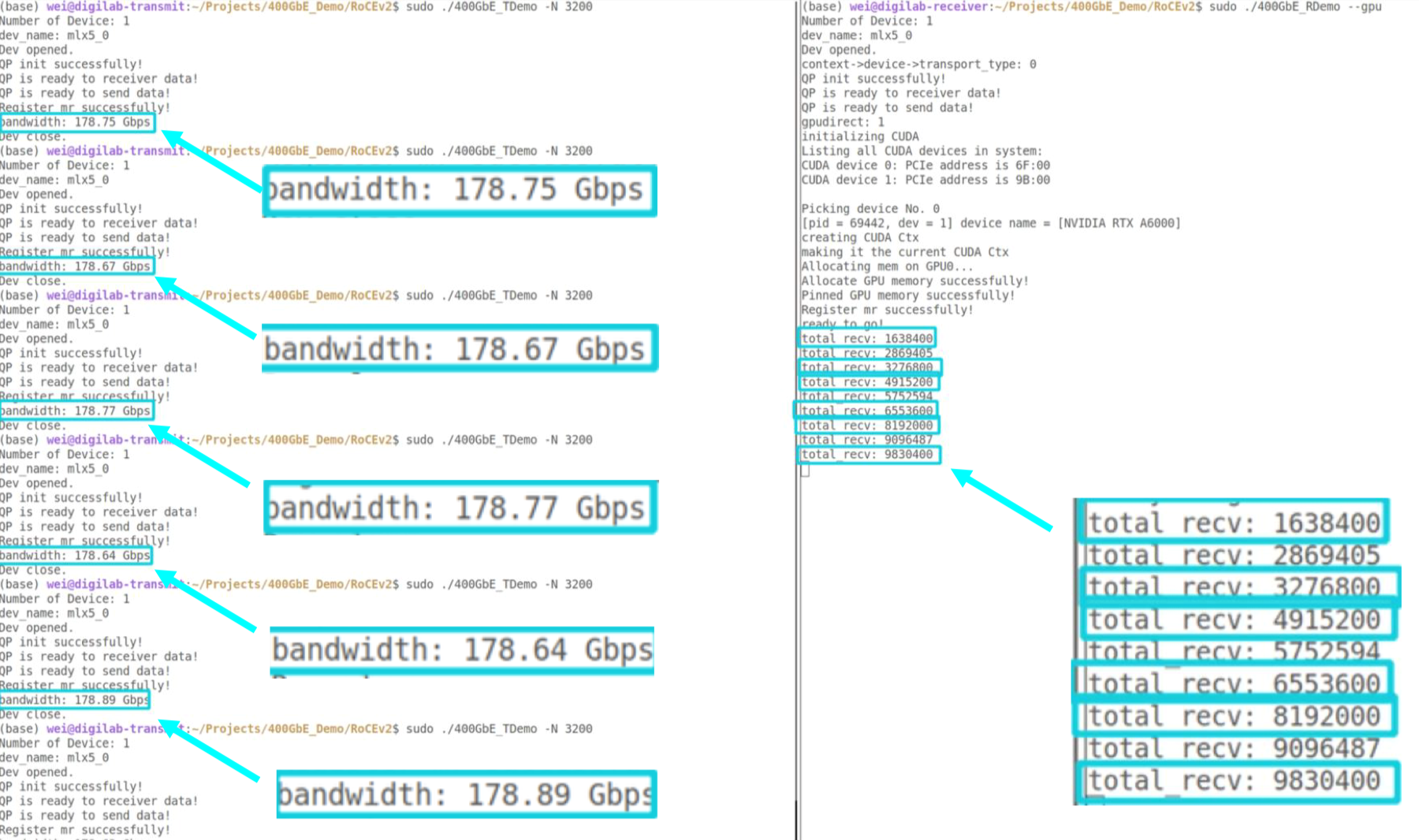}
	\caption{
		Measured bandwidth results of data transfers using GPUDirect to a single GPU on a 200~Gbps link. The maximum data rate achieved without packet loss is 178~Gbps. The packet loss is checked By comparing the number of sent packets and received packets. In each test, 1638400 packets are sent and 1638400 packets are received.
		\label{gpudirect-test-result}
	}
\end{figure}

\newpage
\subsection{Direct data transport from NIC to two GPUs}
\label{section-400g-gpudirect-nic-2gpu}
As the 400G NIC is capable of supporting two 200Gbps links, and each GPU supports 200Gbps IO bandwidth, the capability of streaming to two GPUs simultaneously using GPUDirect was also tested. Figure~\ref{nic-to-2gpu} shows the direct data transfer path from NIC to two GPUs using GPUDirect.\\

\begin{figure}[htbp]
	\centering
	\includegraphics[height=4.5cm]{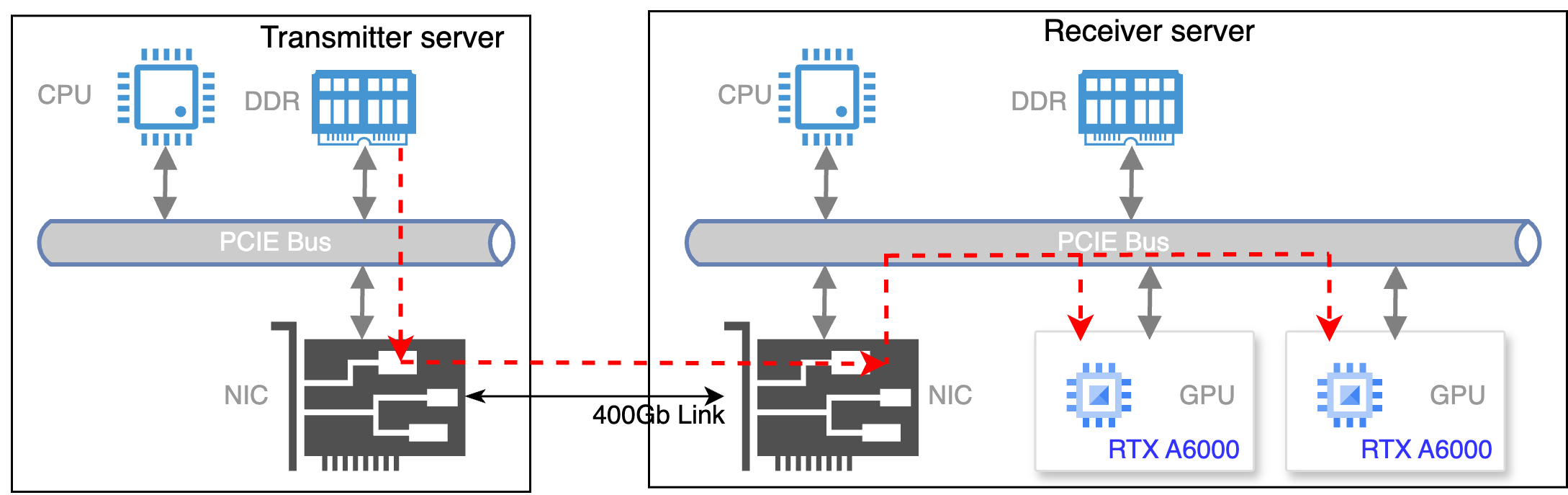}
	\caption{
		Direct data transfer path from NIC to two GPUs using GPUDirect technology. The link between the two servers is 400G link. Red dashed lines show the direction of data transfer.
		\label{nic-to-2gpu}
	}
\end{figure}

On the transmitter server side, the same code based on ibverbs API mentioned in section~\ref{section-400g-gpudirect} was used. In this case, however, the packets contain different source and destination port numbers. On the receiver server side, two receiving queue pairs with different flow are  set for steering the two packet streams to the two GPUs. The resulting throughput ($\sim$180~Gbps)is shown in Figure~\ref{nic-to-2gpu-result}.\\

 \begin{figure}[htbp]
	\centering
	\includegraphics[height=6cm]{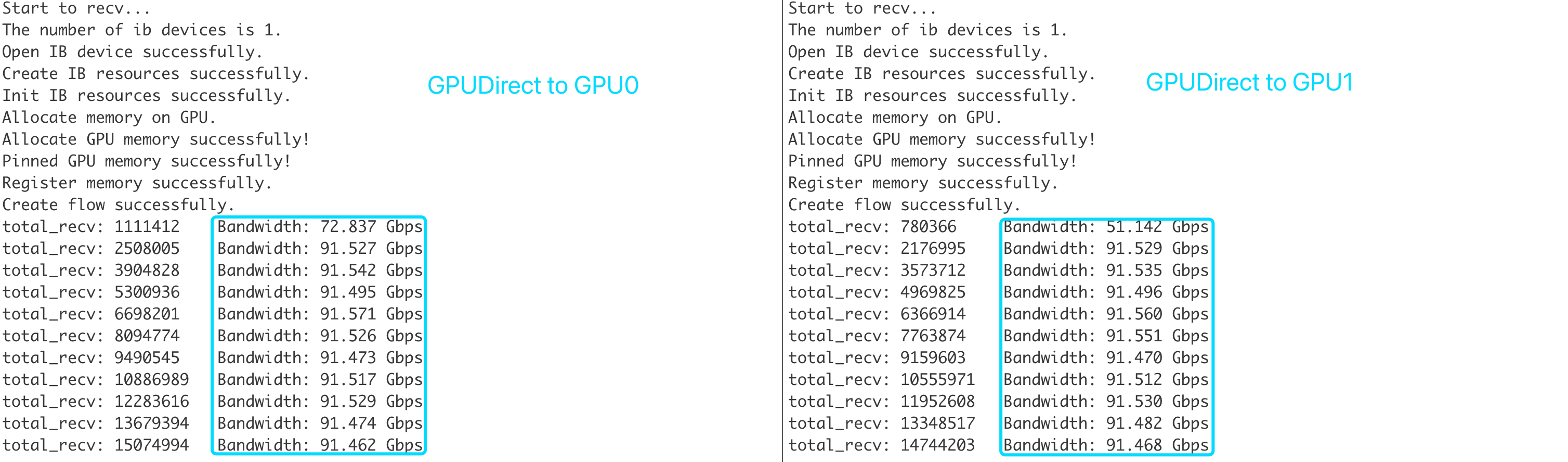}
	\caption{
		Measured bandwidth results of data transfers to two GPUs using GPUDirect on a 400Gbps link. The maximum data rate achieved is $\sim$90Gbps $\times$ 2. 
		\label{nic-to-2gpu-result}
	}
\end{figure}

\newpage
\subsection{Direct Data Transport from two NICs to two GPUs}
\label{section-400g-gpudirect-2nic-2gpu}
While GPUDirect is based entirely on the PCIe bus similarly to RDMA, the performance was more limited, particularly when working with multiple GPUs. Besides measuring the performance of GPUDirect from one NIC to two GPUs, the performance of two pairs of NICs to GPUs with GPUDirect was also tested. The data transfer path is shown in Figure~\ref{2nic-to-2gpu}, and the result ($\sim$276~Gbps) is shown in Figure~\ref{2nic-to-2gpu-result}. The improvement in this rate compared to the single 400Gbps NIC suggests that the limitations in data rate transfer of the single NIC to two GPUs are due to the GPUDirect configuration and PCIe lanes available in the case of a single NIC.\\

\begin{figure}[htbp]
	\centering
	\includegraphics[height=4.5cm]{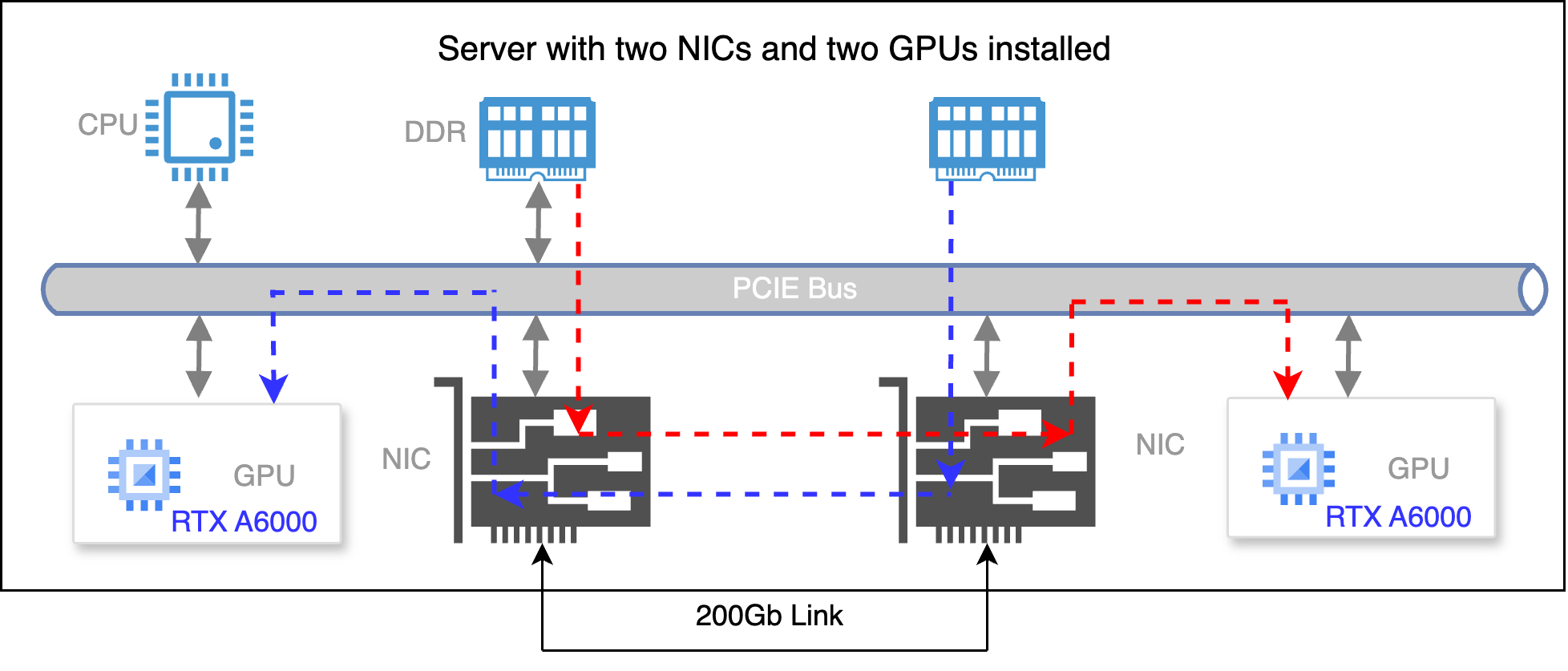}
	\caption{
		Direct data transfer path from two NIC to two GPUs on one server. Both of the two NICs send data to each other, and use GPUDirect to move the received data to two GPUs. The link is 200G. Red and blue dashed lines shows the direction of data transfer.
		\label{2nic-to-2gpu}
	}
\end{figure}

 \begin{figure}[htbp]
	\centering
	\includegraphics[height=9cm]{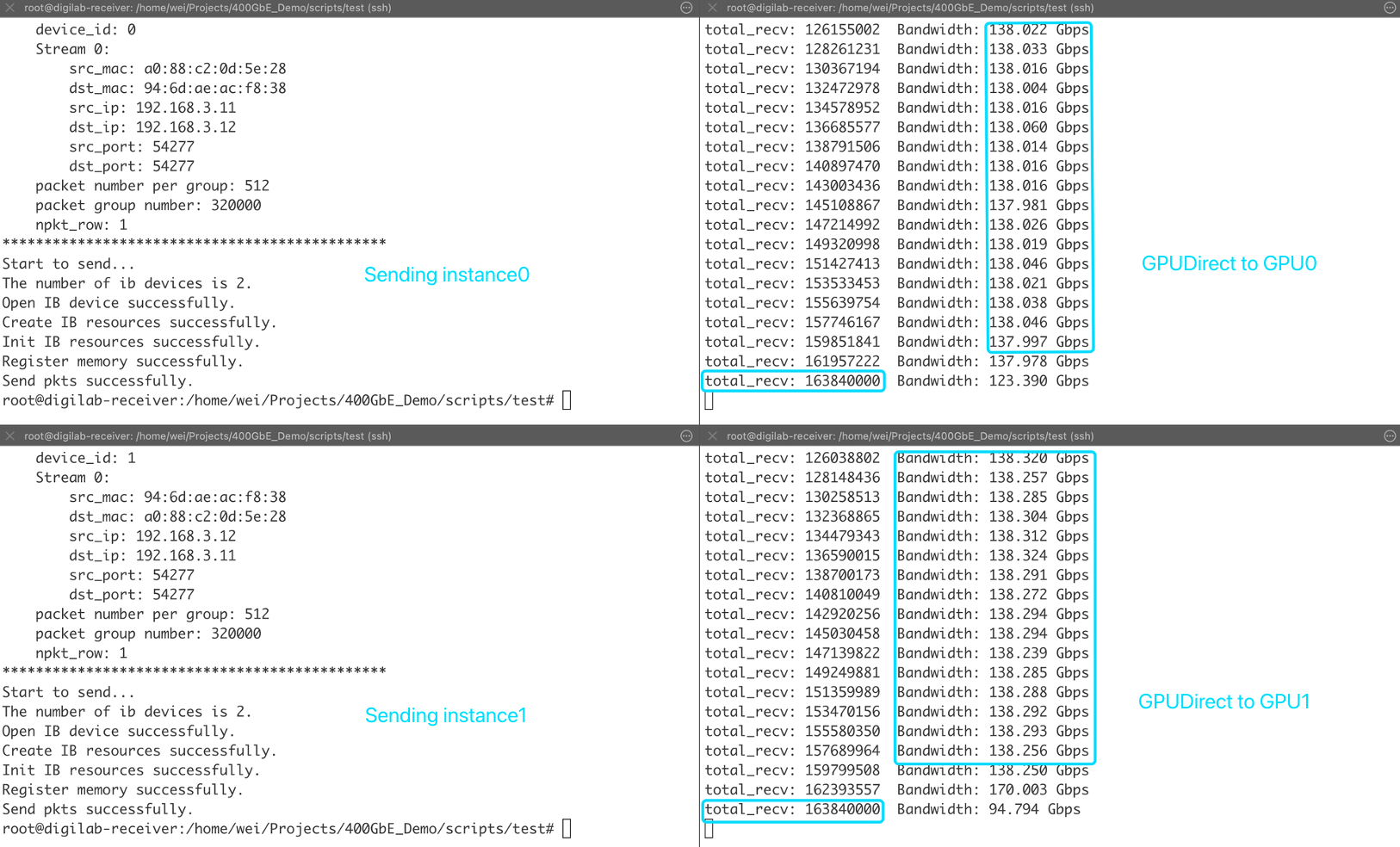}
	\caption{
		Measured bandwidth results of data transfers from two NICs to two GPUs using GPUDirect on a 200Gbps($\times$ 2) link. The maximum data rate achieved is $\sim$138Gbps $\times$ 2. 
		\label{2nic-to-2gpu-result}
	}
\end{figure}

\clearpage
\newpage

\section{400GbE FPGA Core}
\label{section-400gbe-core-imp}
The open source 400GbE FPGA core comprises two primary modules: the UDP packet generation core and the MAC/PHY core. The UDP packet generation core creates the UDP data packet and inserts the user-defined MAC address, IP address, port numbers into the packets. The MAC/PHY core, compliant with the IEEE 802.3ck-2022 400Gbps Ethernet standard, interfaces the FPGA with other devices over copper or optical transceiver modules. The AXI4 bus is used to transport control signals and statistical data. Figure~\ref{400G-FPGA-core} illustrates the core modules.\\

\begin{figure}[htbp]
	\centering
	\includegraphics[height=4.5cm]{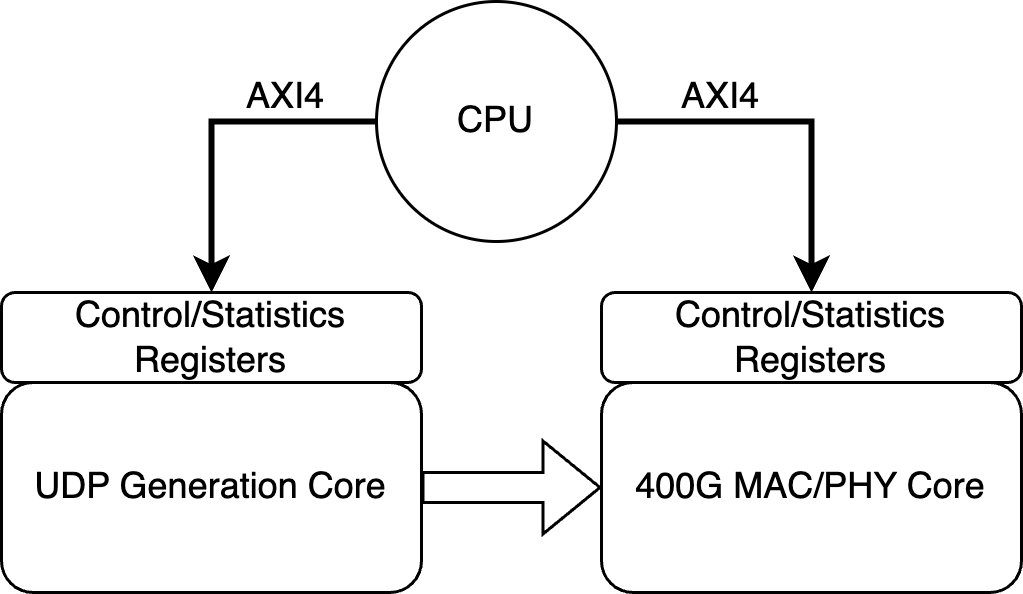}
	\caption{
		Block diagram of the 400GbE FPGA core shows the UDP packet generation and MAC/PHY core modules. 
		\label{400G-FPGA-core}
	}
\end{figure}

\subsection{400GbE Core Implementation}
\subsubsection{UDP packet generation core}
\label{section-400g-udp-gen-imp}
The 400GbE UDP generation core, modified from the CASPER 100G core\footnote{\url{https://github.com/casper-astro/kutleng_skarab2_bsp_firmware}}, consists of two submodules: a streaming data module, and a CPU data module. \\

The streaming data module handles high-speed data transmission and reception to and from the 400G MAC/PHY core. It obtains MAC, IP, and port information via the AXI4 bus to produce packets at a high data rate, stored in registers or the ARP cache. For the TX data path, the module retrieves data from registers or an internal ARP cache, creates TX packets, and sends them to a TX ring buffer. For the RX data path, received packets are stored in an RX ring buffer and filtered by an RX filter to remove any unexpected packets with incorrect MAC, IP, or port numbers. The block diagram of the streaming data module is shown in Figure~\ref{400g-streaming-data-module}. \\

\begin{figure}[htbp]
	\centering
	\includegraphics[height=5.6cm]{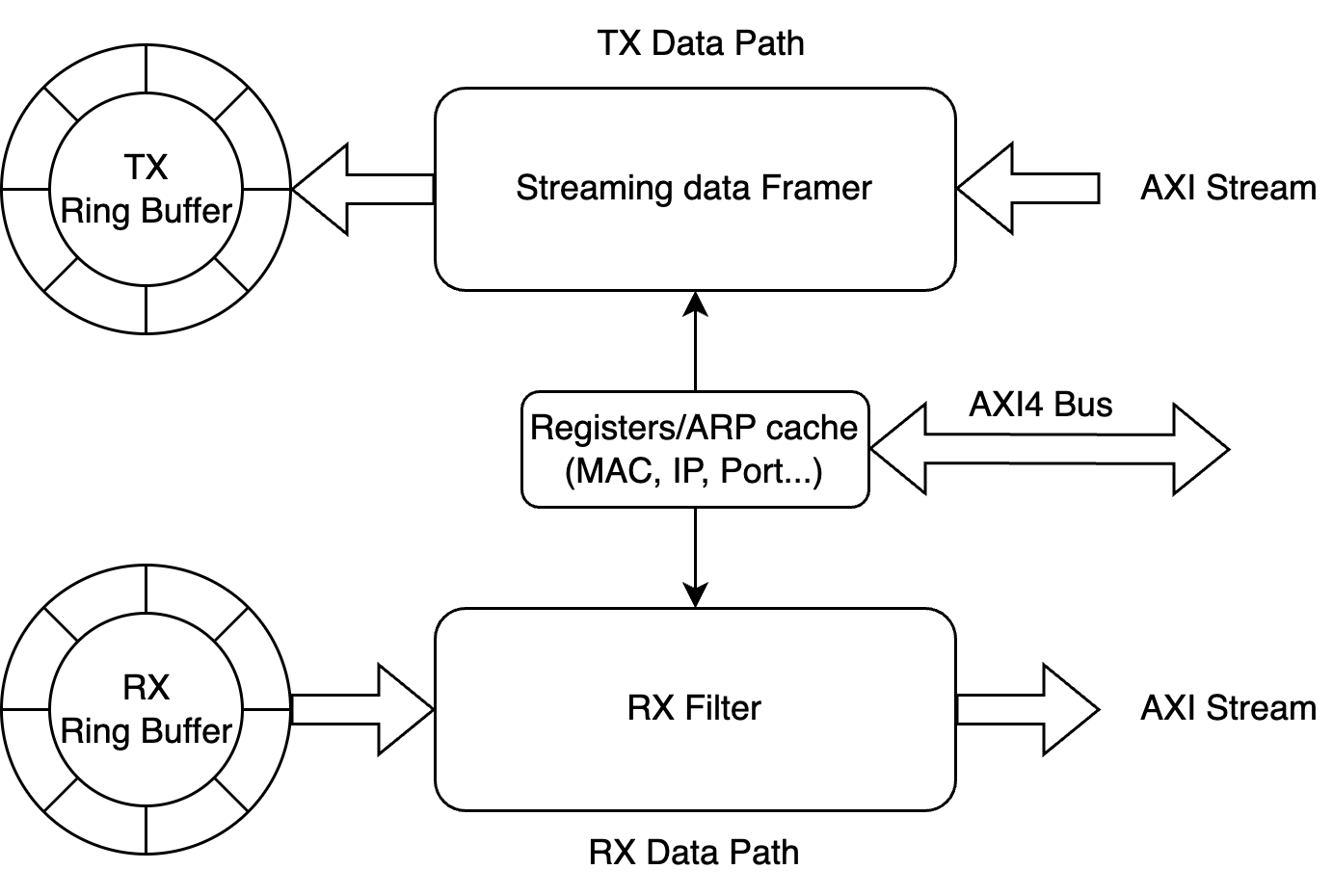}
	\caption{
		Block diagram of the streaming data module within the 400GbE FPGA core. The module handles high-speed data transmission and reception, including packet generation and filtering.
		\label{400g-streaming-data-module}
	}
\end{figure}

The CPU data module sends and receives data to and from a microcontroller implemented in the FPGA. It shares registers, ARP cache, and the MAC/PHY core with the streaming module. An arbitration module determines which data to forward to the MAC/PHY core.  \\

\subsubsection{MAC/PHY Core}
\label{section-400G-mac-phy-core-imp}
In the 400GbE FPGA core, the MAC/PHY core includes the Physical Coding Sublayer (PCS), Physical Medium Attachment (PMA), Physical Medium Dependent (PMD) layer, and support for varying numbers of lanes for various 400GbE protocols.\\

We employ AMD's (Xilinx) GTM transceivers and 600G channelized multi-rate ethernet subsystem (DCMAC) core. The DCMAC is a high-performance, flexible, Ethernet-integrated hard IP that targets various networking applications. The 400GbE, 200GbE, and 100GbE combinations that the core supports allow for a maximum data rate of 600Gbps. For 400GbE, 200GbE, and 100GbE, it implements all of the 400G PCS operations, including: encoder, scrambler, alignment marker insertion, and forward error correction (FEC). Additionally, it partially performs the PMA function, freeing up 8 data lanes (two PCS lanes are consolidated into one) for GTM transceivers to connect to. For our open source 400GbE core design, the DCMAC core is a good choice because the hard IP on Versal SoC is supported by AMD with a free license. \\

With capabilities for up to 56Gbps per lane, the Versal SoC's GTM transceiver is the highest performing AMD transceiver. Working in full-density mode or half-density mode, it carries out the remaining PMA functions. For 400GbE applications, the interface is 400GAUI-8 in full-density mode since each GTM transceiver operates independently; in half-density mode, two GTM transceivers cooperate to support 112Gbps per lane. In this scenario, the interface is 400GAUI-4. Since the NIC in our application supports 400GBASE-CR4, in order to obtain 106 Gbps/lane $\times$ 4 lanes, we must configure the GTM in half density mode. We use two GTM quads to achieve the 400Gbps data throughput, because each GTM quad contains four GTM transceivers. The MAC/PHY core, based on DCMAC and GTM transceiver, is shown in Figure~\ref{400g-mac-phy-core}. The complete FPGA implementation of the 400GbE architecture is shown in Figure~\ref{400g-whole-system}.

\begin{figure}[htbp]
	\centering
	\includegraphics[height=8cm]{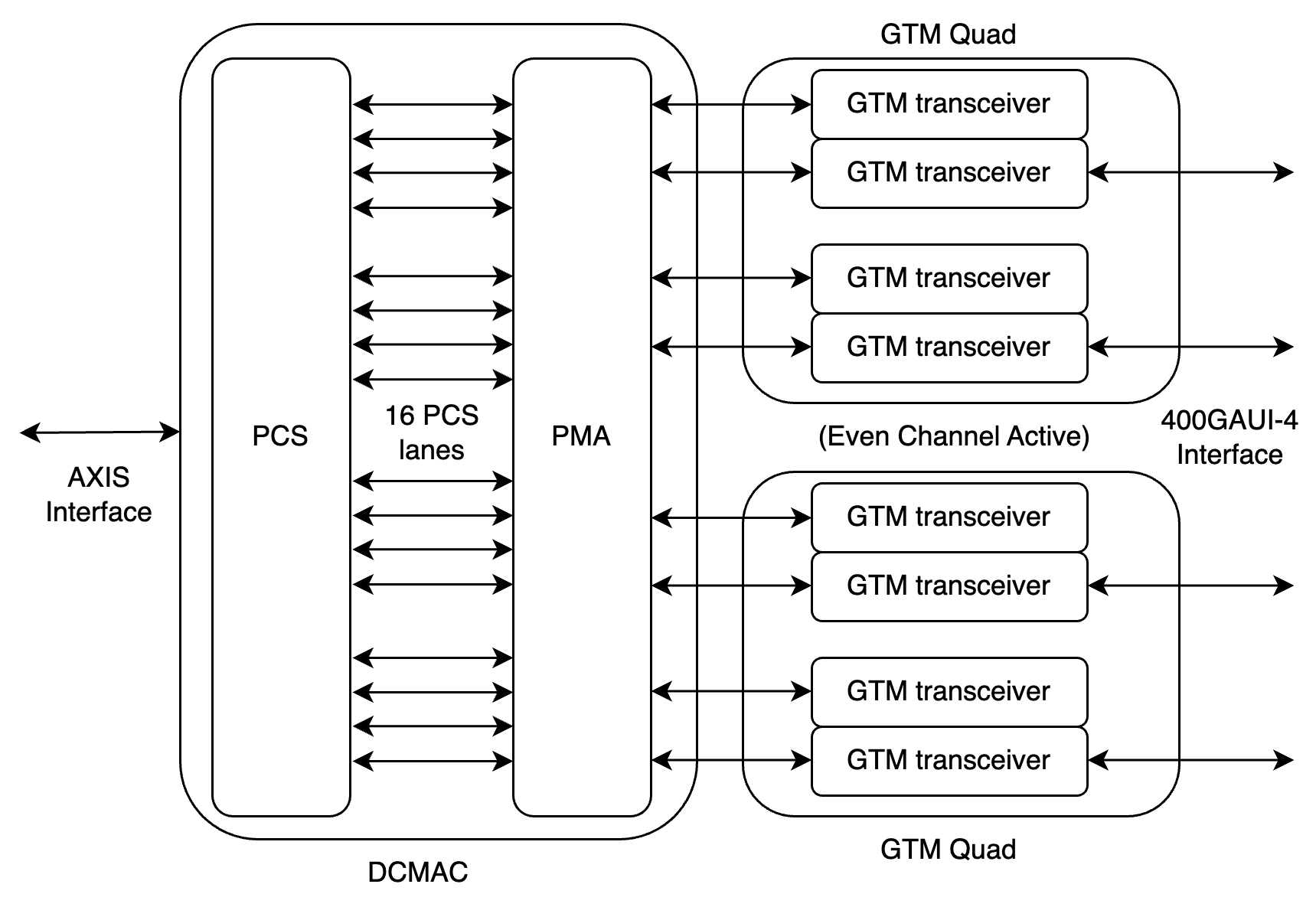}
	\caption{
		Integrated DCMAC core and GTM transceivers in the 400GbE MAC/PHY core. The components work together to support high-speed Ethernet communication.
		\label{400g-mac-phy-core}
	}
\end{figure}

\begin{figure}[htbp]
	\centering
	\includegraphics[height=7cm]{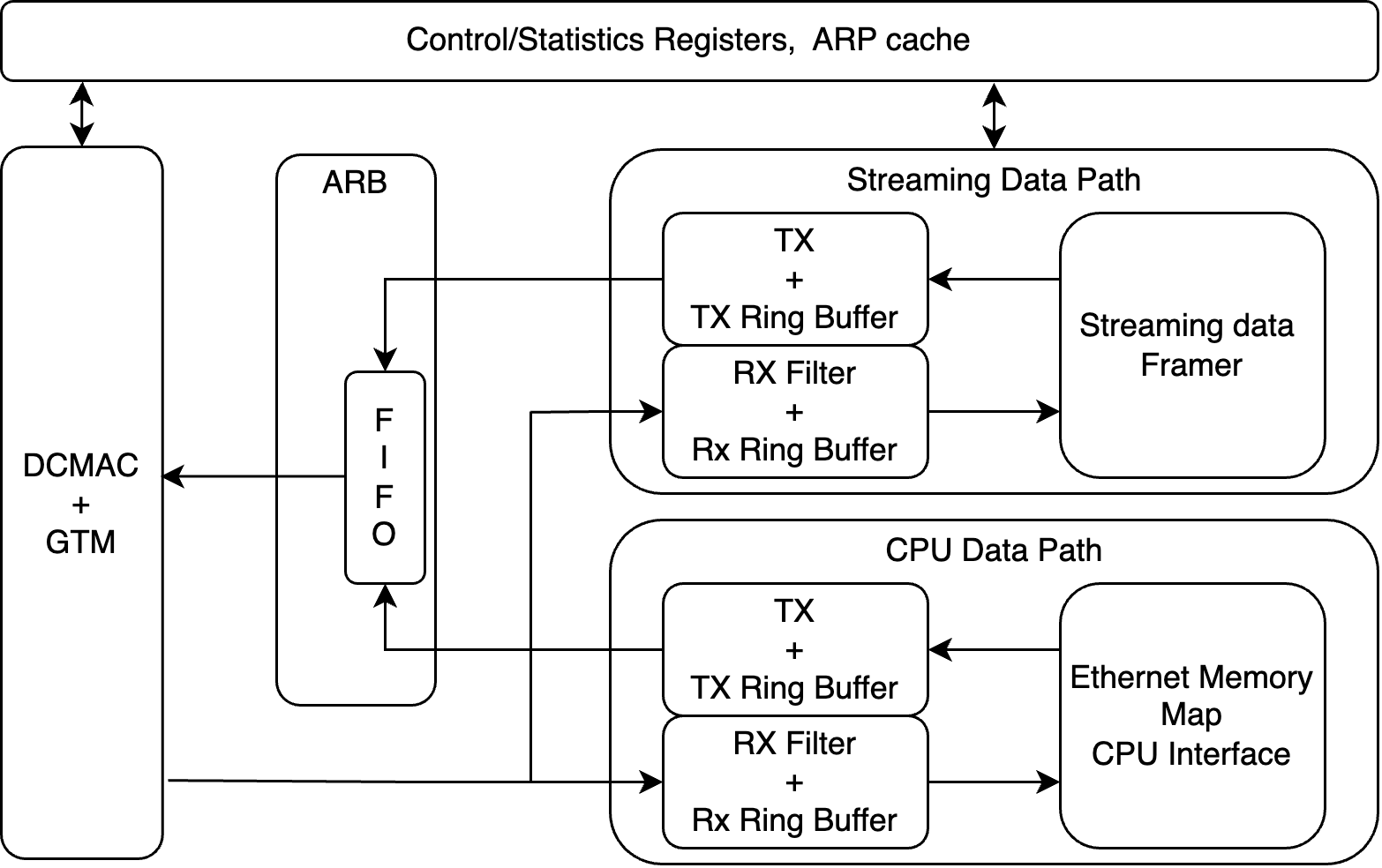}
	\caption{
		Complete architecture of the 400GbE FPGA core, illustrating the integration of various modules to achieve 400Gbps data transfer rates.
		\label{400g-whole-system}
	}
\end{figure}
\newpage
\subsection{FPGA to NIC to GPU through DRAM result}
\label{section-fpga2dram2nic-result}
To evaluate bandwidth performance of the 400GbE FPGA core, we developed a packet generation module that is connected to the 400GbE core. Each packet contains a 16-bit counter for packet loss detection, and the packet rate can be adjusted to obtain the data rate without packet loss. Using the same HASHPIPE framework-based code as described in Section~\ref{section-400g-nic-dram-gpu}, we monitor the packet loss status and real-time bandwidth measurement. Figure~\ref{400g-fpga-nic-diagram} illustrates the test setup, and Figure~\ref{400g-fpga-nic-setup} shows the actual configuration. 

\begin{figure}[htbp]
	\centering
	\includegraphics[height=4.5cm]{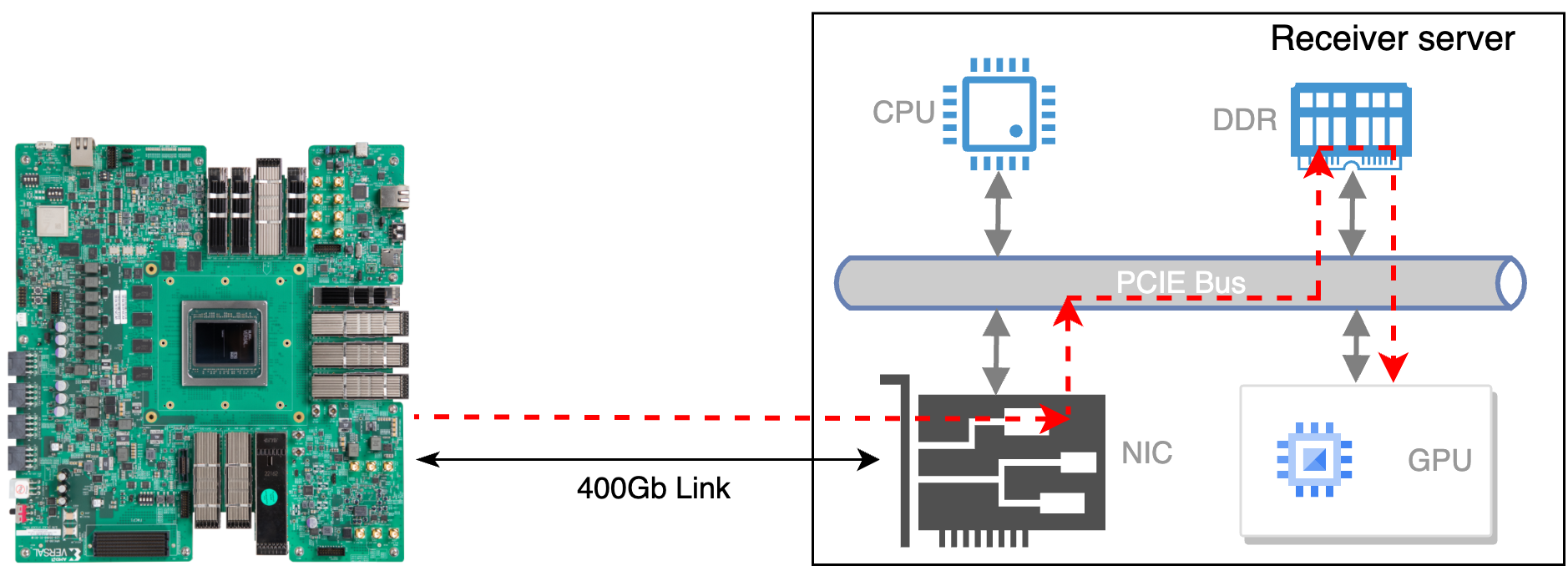}
	\caption{
		Diagram showing the setup for the FPGA to NIC to GPU via DRAM data transfer test. The setup includes the FPGA board, 400G NIC, and GPUs to evaluate data transfer performance.
		\label{400g-fpga-nic-diagram}
	}
\end{figure}

\begin{figure}[htbp]
	\centering
	\includegraphics[height=5cm]{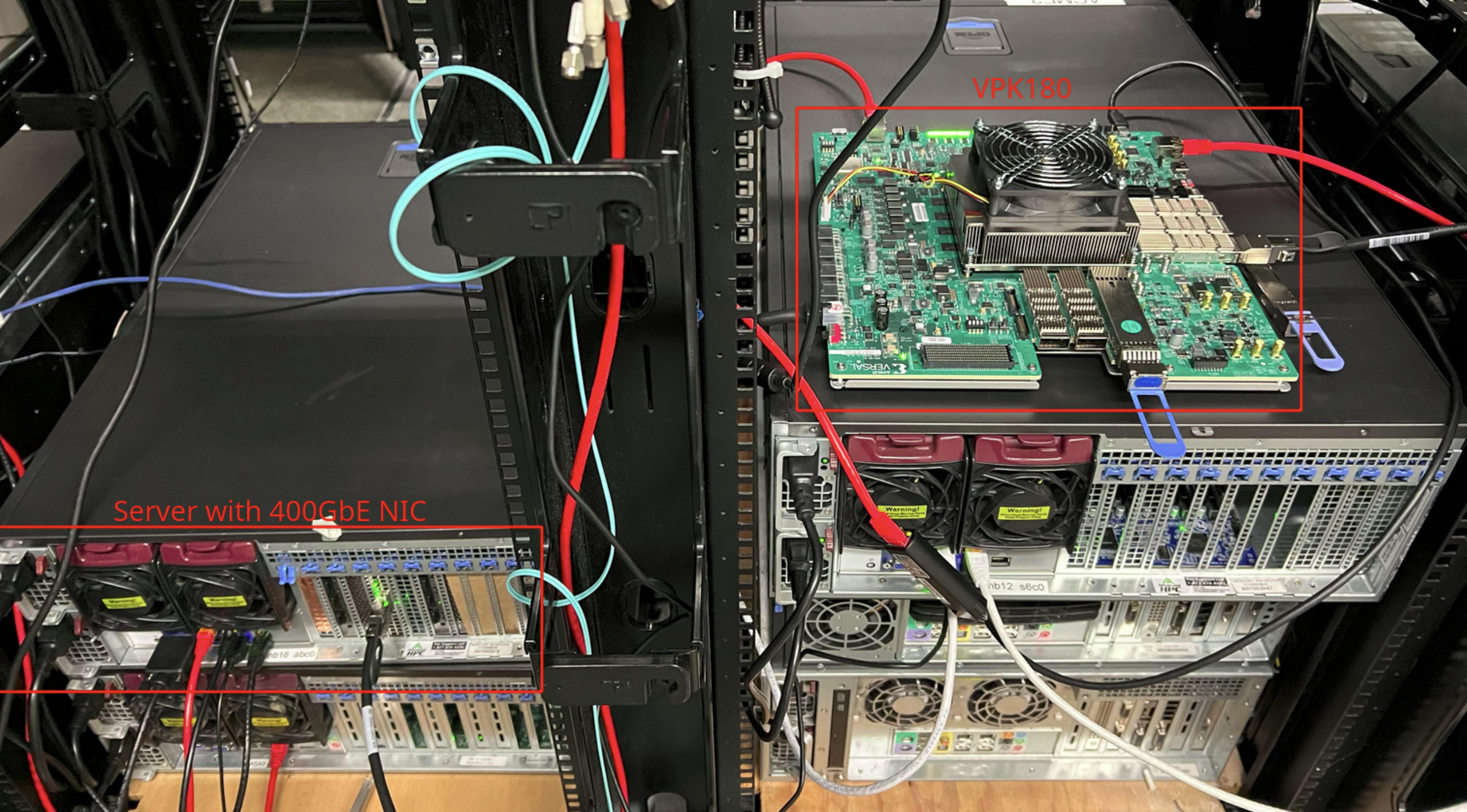}
	\caption{
		The actual setup used for the FPGA to NIC to GPU test. The setup demonstrates the hardware configuration and connections used in the experiment.
		\label{400g-fpga-nic-setup}
	}
\end{figure}

Before further testing, we confirmed the 400G link between the FPGA and the server with a 400G NIC. Once configured with the 400GbE firmware, the server indicated an active state at 400G with four active lanes. Figure~\ref{400g-fpga-linkup} shows the receiver NIC state.\\

\begin{figure}[htbp]
	\centering
	\includegraphics[height=9cm]{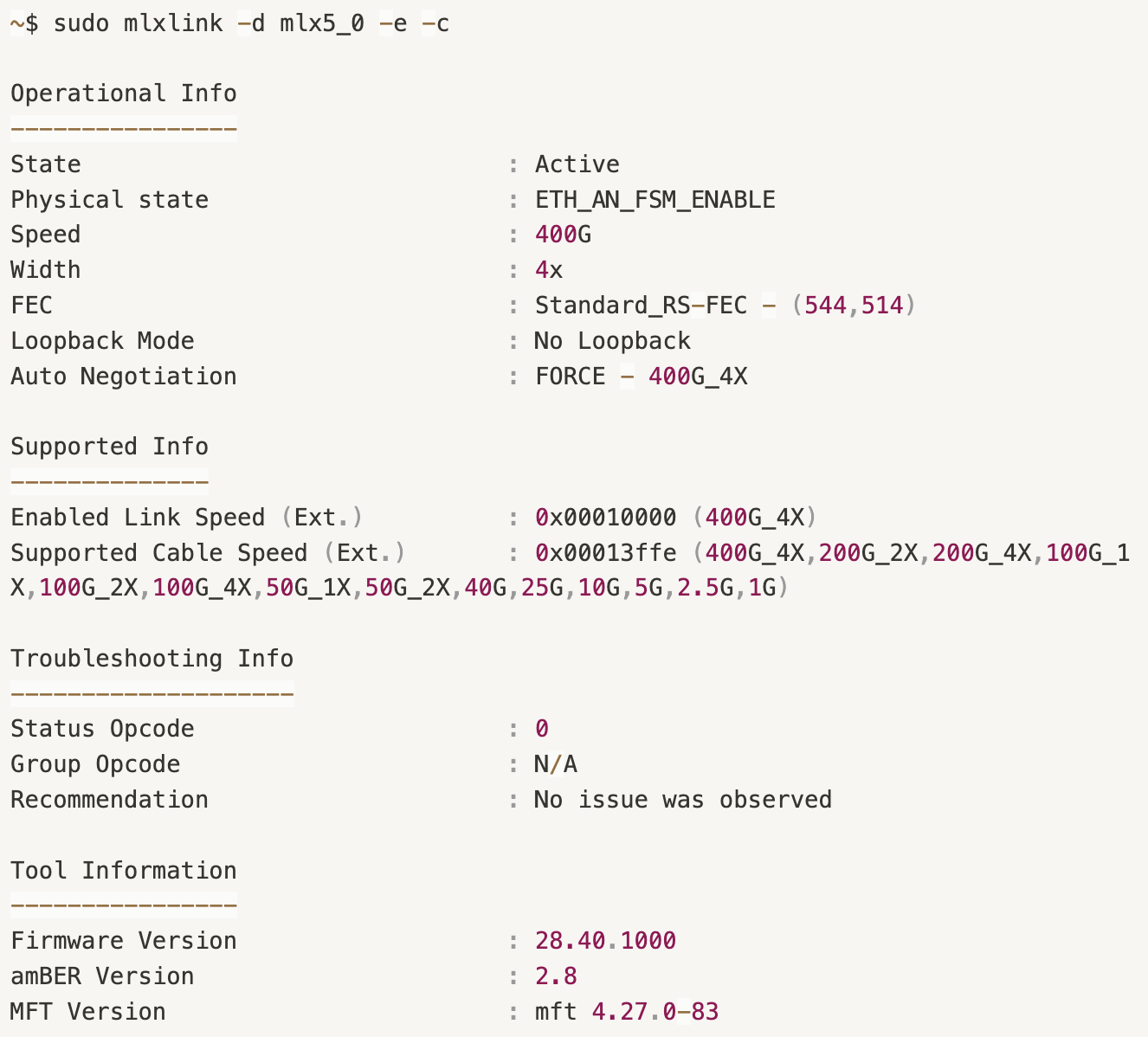}
	\caption{
		Status indication of the FPGA linked to the 400G NIC at a 400~Gbps data rate. The figure confirms the active state and link speed of the connection.
		\label{400g-fpga-linkup}
	}
\end{figure}

The GPUs used in this test are two RTX A6000 and two RTX 4070. The clock rate on the FPGA side for generating packets and clocking the 400G Ethernet core is 390.625MHz, and the bus width of the DCMAC core is 1024 bits, so the total data rate from the DCMAC core is up to 400Gbps. After adjusting the speed of generating packets, the code running on the server side can capture the packets at up to $\sim$362~Gbps without packet loss. We observed no packet loss with data rates up to 362~Gbps. This result is identical as we obtained in Section~\ref{section-400g-nic-dram-gpu}. Figure~\ref{fpga-nic-dram-gpu-result} shows the results of the FPGA to NIC to two GPUs test.\\

\begin{figure}[htbp]	
	\subfigure[RTX A6000 FPGA to GPU bandwidth performance]
	{
			\label{a6000-fpga-nic-dram-gpu}
			\includegraphics[height=3.4cm]{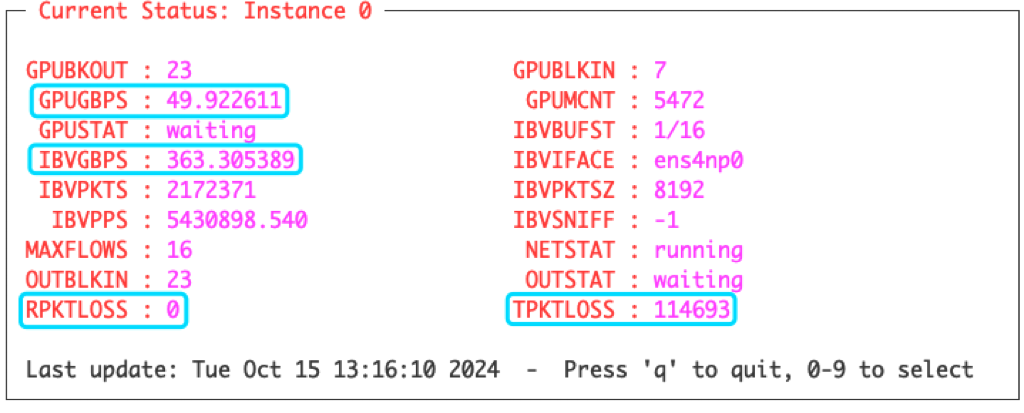}	}
	\subfigure[RTX 4070 FPGA to GPU bandwidth performance]
	{
			\label{4070-fpga-nic-dram-gpu}
			\includegraphics[height=3.4cm]{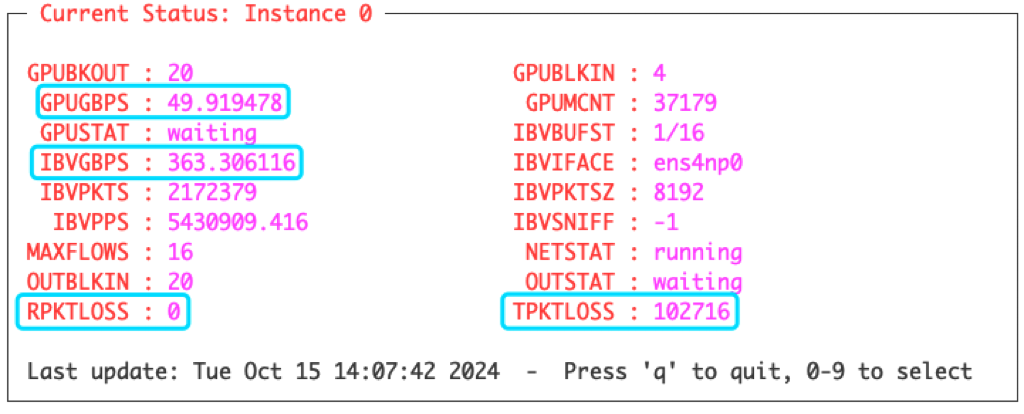}   
	}
	\caption{Test results of the FPGA to NIC to two GPUs data transfer, showing a data rate of 362Gbps without packet loss. RPKTLOSS shows the real-time packet loss, which is always zero after the initialization; TPKTLOSS shows the total number of packet loss, which is non-zero during the initialization, but remains constant after initialization. The figure illustrates the effectiveness of the 400GbE FPGA core in high-speed data transfer. } 
	\label{fpga-nic-dram-gpu-result}
\end{figure}

\newpage

\subsection{FPGA to GPU Test Result with GPUDirect}
\label{section-fpga-gpudirect}
As the GPUs we are using in the tests have PCIe~4.0 interface with the speed limitation to $\sim$200Gbps, what we expected is moving data into one GPU at $\sim$200Gbps. To send packets from FPGA to two GPUs with GPUDirect technology, packets are sent out from FPGA with different source and dest port numbers, and these two kinds of packets are sent out one by one. On the server side, two queue pairs with differet steer flows are created for directing the two kinds of packets into two GPUs. Figure~\ref{fpga-gpu-gpudirect} shows the test about FPGA to two GPUs with GPUDirect.\\

\begin{figure}[htbp]
	\centering
	\includegraphics[height=4.5cm]{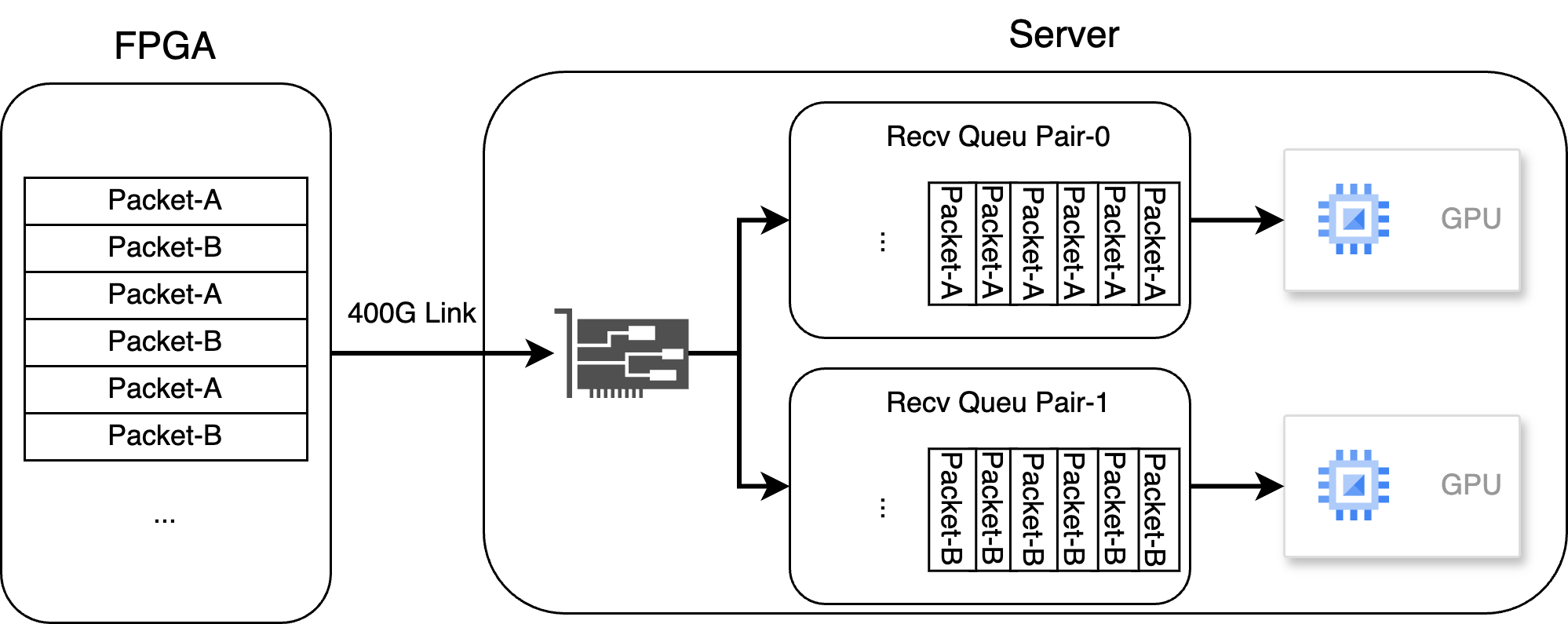}
	\caption{
		FPGA and server are connected with 400G link. FPGA sends out two kinds of packets with different source and dest port numbers: Packet-A and Packet-B. The data stream are steered to two queue pairs with different steer flow, and then sent to two GPUs directly. 
		\label{fpga-gpu-gpudirect}
	}
\end{figure}
If only one receiving queue pair is enabled, half of the data will be sent to one GPU. The bandwidth is about $\sim$180Gbps, which is identical to the result mentioned in section\ref{section-400g-gpudirect}; If both of the two receiving queue pairs are enable, the one data stream will be split to two data streams, and the data will be sent to two GPUs at the same time directly. The bandwidth of each data stream is $\sim$90Gbps, and the total bandwidth is $\sim$180Gbps. The results about this test are shown in Figure~\ref{fpga-gpu-gpudirect-1qp} and figure\ref{fpga-gpu-gpudirect-2qp}.

\begin{figure}[htbp]	
	\centering
	\subfigure[GPUDirect from FPGA to GPU with one queue pair enabled]
	{
			\label{fpga-gpu-gpudirect-1qp}
			\includegraphics[height=8.3cm]{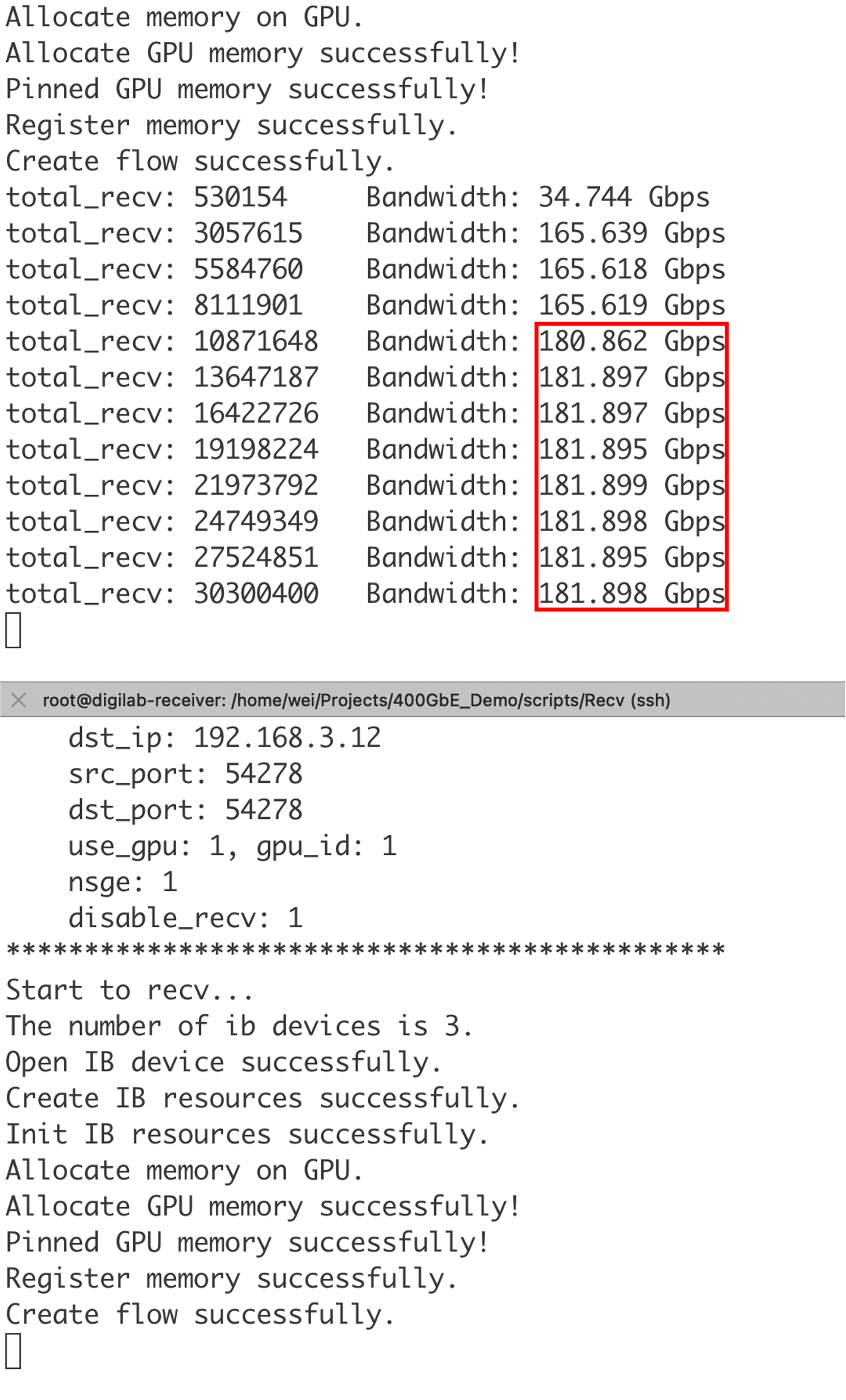}	}
	\subfigure[GPUDirect from FPGA to GPU with two queue pairs enabled]
	{
			\label{fpga-gpu-gpudirect-2qp}
			\includegraphics[height=8.1cm]{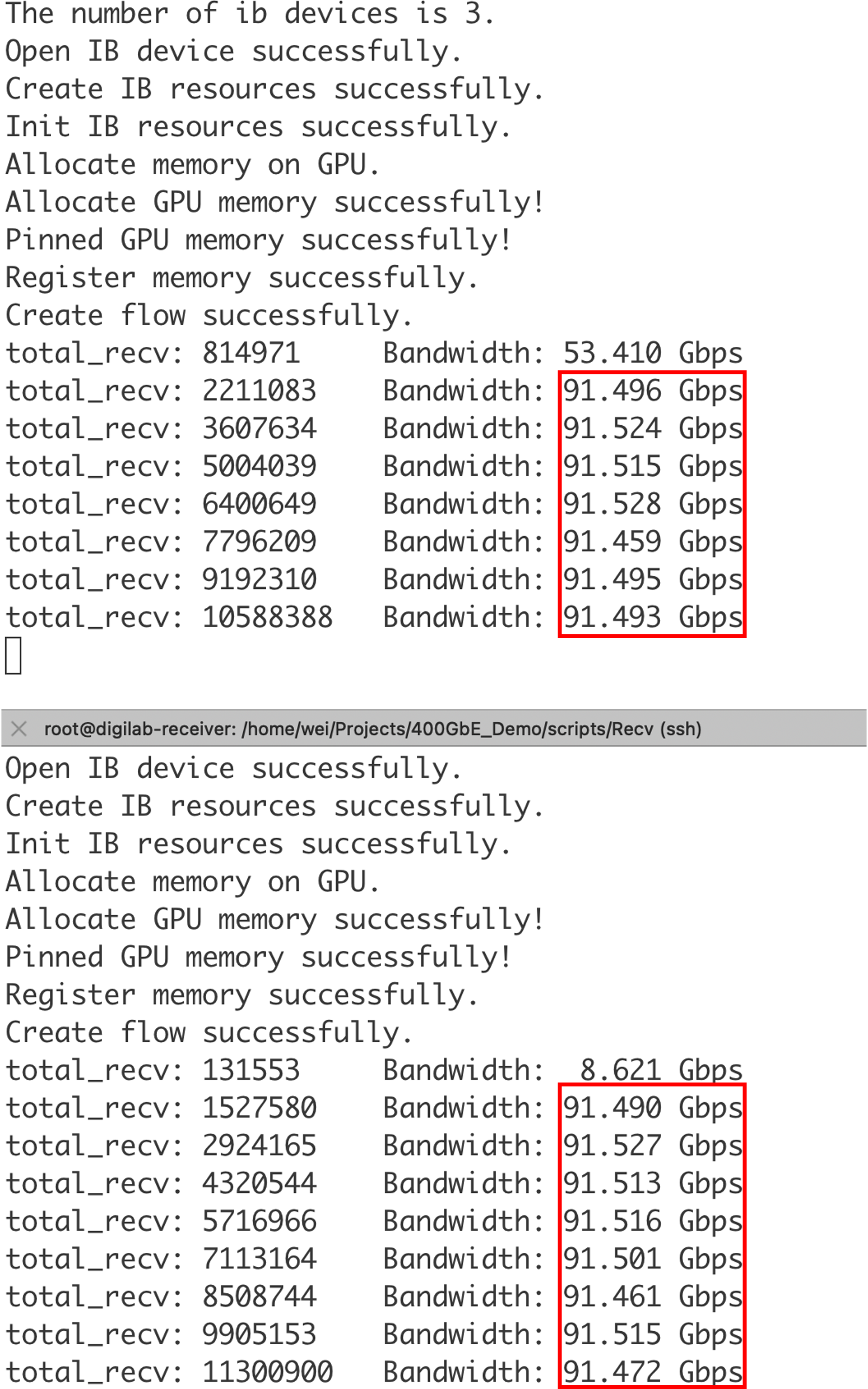}   
	}
	\caption{GPUDirect from FPGA to NIC results. If one queue pair is enabled on the receiver side, the bandwidth is about $\sim$180Gbps; if both of the two queue pairs are enabled, bandwidth for each data stream is $\sim$90Gbps, and total is about $\sim$180Gbps.} 
\end{figure}

\newpage

\section{Summary and Future Work}
\label{section-summary}
This study examines the bandwidth performance of two RDMA techniques: moving data through DRAM and GPUDirect, when transferring data from NIC to GPUs. Moving data to GPUs through DRAM can achieve approximately 180~Gbps to a single GPU without packet loss and about 360~Gbps to two GPUs. With GPUDirect, we can transfer $\sim$180Gbps to a single GPU, and $\sim$90Gbps $\times$ 2 to two GPUs simultaneously. Using two NICs and two GPUs with GPUDirect can achieve the bandwidth of $\sim$138Gbps $\times$ 2.  When using GPUDirect to two GPUs from a single NIC, the data rate was half of that expected. In the case of the dual NICs to dual GPUs the data rate improved, but was still below that achieved when transferring the data via DRAM. We plan to conduct research on further high-performance computing methods in the future, including Data Plane Development Kit (DPDK) and Holoscan in order to compare the performance of these approaches to our current results.\\

In addition to the server to server tests, a 400GbE FPGA core was developed. As many radio telescopes using a networked architecture, this allowed the testing of a typical radio astronomy scenario where simple UDP packets were transmitted to a server. Using this core, data rates of 362~Gbps were transferred without experiencing any packet loss between the FPGA and two GPUs. Testing with GPUDirect, the 400GbE FPGA core can transfer data at about 180~Gbps to one GPU. This core will be added to the CASPER library, and the test results demonstrate the potential of RDMA techniques and the FPGA 400GbE core for high data rate applications.\\

\section{Acknowledgement}
\label{section-acknowledgement}
The authors would like to thank John Romein (ASTRON) for his advice on 400G NIC settings, Andre Renard (UofT)) for his advice on memory bandwidth improvement, Adam Thompson and Cliff Burdick (NVIDIA) for their assistance with the NIC to NIC tests, and Leo Karnan (AMD) for his advice on GTM transceiver settings. The authors also thank to AMD's University Program for their donation of the VPK180 FGPA board. The research was funded by NSF grants 2009537 and 2307781.


\bibliography{sample.bib}

\begin{thebibliography}{}
\expandafter\ifx\csname natexlab\endcsname\relax\def\natexlab#1{#1}\fi
\providecommand{\url}[1]{\href{#1}{#1}}
\providecommand{\dodoi}[1]{doi:~\href{http://doi.org/#1}{\nolinkurl{#1}}}
\providecommand{\doeprint}[1]{\href{http://ascl.net/#1}{\nolinkurl{http://ascl.net/#1}}}
\providecommand{\doarXiv}[1]{\href{https://arxiv.org/abs/#1}{\nolinkurl{https://arxiv.org/abs/#1}}}

\bibitem[{Ad{\'a}mek {et~al.}(2021)Ad{\'a}mek, Novotn{\`y}, Thiyagalingam, \& Armour}]{gpufft2021}
Ad{\'a}mek, K., Novotn{\`y}, J., Thiyagalingam, J., \& Armour, W. 2021, IEEE Access, 9, 18167

\bibitem[{Agarwal {et~al.}(2020)Agarwal, Lorimer, Surnis, Pei, Karastergiou, Golpayegani, Werthimer, Cobb, McLaughlin, White, {et~al.}}]{gpufrb2020}
Agarwal, D., Lorimer, D., Surnis, M., {et~al.} 2020, Monthly Notices of the Royal Astronomical Society, 497, 352

\bibitem[{Akeret {et~al.}(2017)Akeret, Chang, Lucchi, \& Refregier}]{gpurfi2017}
Akeret, J., Chang, C., Lucchi, A., \& Refregier, A. 2017, Astronomy and computing, 18, 35

\bibitem[{Alvear {et~al.}(2016)Alvear, Finger, Fuentes, Sapunar, Geelen, Curotto, Rodr{\'\i}guez, Monasterio, Reyes, Mena, {et~al.}}]{fpga2016}
Alvear, A., Finger, R., Fuentes, R., {et~al.} 2016, in Millimeter, Submillimeter, and Far-Infrared Detectors and Instrumentation for Astronomy VIII, Vol. 9914, SPIE, 332--345

\bibitem[{Ayala {et~al.}(2022)Ayala, Tomov, Luszczek, Cayrols, Ragghianti, \& Dongarra}]{fft-compare}
Ayala, A., Tomov, S., Luszczek, P., {et~al.} 2022, Univ. Tennessee at Knoxville, Knoxville, TN, USA, Tech. Rep. ICL-UT-22--02

\bibitem[{Brown {et~al.}(2004)Brown, Wild, \& Cunningham}]{alma2004}
Brown, R.~L., Wild, W., \& Cunningham, C. 2004, Advances in Space Research, 34, 555

\bibitem[{Carlson(2000)}]{widar}
Carlson, B. 2000, Implementation, and Signal Processing, NRC-EVLA Memo, 1

\bibitem[{Clark {et~al.}(2013)Clark, Plante, \& Greenhill}]{xgpu2012}
Clark, M.~A., Plante, P.~L., \& Greenhill, L.~J. 2013, The International journal of high performance computing applications, 27, 178

\bibitem[{Connor {et~al.}(2022)Connor, Bouman, Ravi, \& Hallinan}]{dsa20002022}
Connor, L., Bouman, K.~L., Ravi, V., \& Hallinan, G. 2022, Monthly Notices of the Royal Astronomical Society, 514, 2614

\bibitem[{Dewdney {et~al.}(2009)Dewdney, Hall, Schilizzi, \& Lazio}]{ska2009}
Dewdney, P.~E., Hall, P.~J., Schilizzi, R.~T., \& Lazio, T. J.~L. 2009, Proceedings of the IEEE, 97, 1482

\bibitem[{Hallinan {et~al.}(2019)Hallinan, Ravi, Weinreb, Kocz, Huang, Woody, Lamb, D'Addario, Catha, Shi, {et~al.}}]{hallinan2019dsa}
Hallinan, G., Ravi, V., Weinreb, S., {et~al.} 2019, arXiv preprint arXiv:1907.07648

\bibitem[{Hickish {et~al.}(2016)Hickish, Abdurashidova, Ali, Buch, Chaudhari, Chen, Dexter, Domagalski, Ford, Foster, {et~al.}}]{casper2016}
Hickish, J., Abdurashidova, Z., Ali, Z., {et~al.} 2016, Journal of Astronomical Instrumentation, 5, 1641001

\bibitem[{Jiang {et~al.}(2020)Jiang, Yu, Chen, \& Liu}]{adc2020}
Jiang, H., Yu, C.-Y., Chen, M., \& Liu, M. 2020, Publications of the Astronomical Society of the Pacific, 132, 085001

\bibitem[{{Jonas} \& {MeerKAT Team}(2016)}]{meerkat2016}
{Jonas}, J., \& {MeerKAT Team}. 2016, in MeerKAT Science: On the Pathway to the SKA, 1, \dodoi{10.22323/1.277.0001}

\bibitem[{Kalia {et~al.}(2016)Kalia, Kaminsky, \& Andersen}]{rdma2016}
Kalia, A., Kaminsky, M., \& Andersen, D.~G. 2016, in 2016 USENIX Annual Technical Conference (USENIX ATC 16), 437--450

\bibitem[{Li {et~al.}(2019)Li, Song, Chen, Li, Liu, Tallent, \& Barker}]{nvlink2019}
Li, A., Song, S.~L., Chen, J., {et~al.} 2019, IEEE Transactions on Parallel and Distributed Systems, 31, 94

\bibitem[{Liu {et~al.}(2021)Liu, Meng, Wang, Zhou, Yao, \& Tariq}]{fpga2021}
Liu, W., Meng, Q., Wang, C., {et~al.} 2021, Journal of Instrumentation, 16, P08047

\bibitem[{MacMahon {et~al.}(2018)MacMahon, Price, Lebofsky, Siemion, Croft, DeBoer, Enriquez, Gajjar, Hellbourg, Isaacson, {et~al.}}]{hashpipe2018}
MacMahon, D.~H., Price, D.~C., Lebofsky, M., {et~al.} 2018, Publications of the Astronomical Society of the Pacific, 130, 044502

\bibitem[{Romein(2021)}]{gpucorrelator221}
Romein, J.~W. 2021, Astronomy \& Astrophysics, 656, A52

\bibitem[{Schilizzi {et~al.}(2007)Schilizzi, Alexander, Cordes, Dewdney, Ekers, Faulkner, Gaensler, Hall, Jonas, \& Kellermann}]{ska2007}
Schilizzi, R., Alexander, P., Cordes, J., {et~al.} 2007, SKA Memorandum, 100

\bibitem[{Yu {et~al.}(2024)Yu, Deng, Sun, Niu, Li, Wu, Wang, Wang, Zuo, Shu, {et~al.}}]{gpufrb2024}
Yu, Z., Deng, F., Sun, S., {et~al.} 2024, Research in Astronomy and Astrophysics

\end{thebibliography}
\bibliographystyle{aasjournal}



\end{document}